\documentclass[11pt,a4paper]{article}
\pdfoutput=1
\usepackage{jheppub}
\usepackage{amsfonts,amssymb,amsmath,graphicx}
\usepackage{caption}
\usepackage{subcaption}
\newcommand{\be}{\begin{equation}}
\newcommand{\ee}{\end{equation}}
\newcommand{\beq}{\begin{equation}}
\newcommand{\eeq}{\end{equation}}
\newcommand{\bea}{\begin{eqnarray}}
\newcommand{\eea}{\end{eqnarray}}

\def\s{\sigma}
\def\f{\frac}

\title{Velocity Statistics in Holographic Fluids: Magnetized Quark-Gluon Plasma and Superfluid Flow}
\author[a]{Daniel Are\'an}
\author[b]{Leopoldo A. Pando Zayas}
\author[c]{Leonardo Pati\~no}
\author[c]{ and Mario Villasante}
\affiliation[a]{Max-Planck-Institut f\"ur Physik (Werner-Heisenberg-Institut,  F\"ohringer Ring 6, D-80805, Munich, Germany }
\affiliation[b]{The Abdus Salam International Centre for Theoretical Physics, Strada Costiera 11, 34014 Trieste, Italy}
\affiliation[b]{Michigan Center for Theoretical Physics, Department of Physics, University of Michigan,
450 Church Street, Ann Arbor, MI 48109, U.S.A.}
\affiliation[c]{Departamento de F\'\i sica, Facultad de Ciencias, Universidad Nacional Aut\'onoma de M\'exico, A.P. 70-542, M\'exico D.F. 04510, M\'exico}

\date{\today}

\abstract{We study the velocity statistics distribution of an external heavy particle in holographic fluids. We argue that when the dual supergravity background has a finite temperature horizon the velocity statistics goes generically as $1/v$, compatible with the jet-quenching intuition from the quark-gluon plasma. A careful analysis of the behavior of the classical string whose apparent worldsheet horizon deviates from the background horizon reveals that other regimes are possible. We numerically discuss two cases: the magnetized quark-gluon plasma and a model of superfluid flow.  We explore a range of parameters in these top-down supergravity solutions including, respectively, the magnetic field and the superfluid velocity.  We determine that the velocity statistics goes largely as $1/v$, however, as we leave the non-relativistic regime we observe some deviations. }


\emailAdd{darean@mpp.mpg.de}
\emailAdd{lpandoz@umich.edu}
\emailAdd{leopj@ciencias.unam.mx}
\emailAdd{mario.villasante@ciencias.unam.mx}

\begin{document}

\begin{flushright}
MCTP-16-13\\
MPP-2016-118
\end{flushright}

\maketitle
\setlength{\parskip}{8pt}


\section{Introduction}

One of the most remarkable paradigms emerging from the AdS/CFT correspondence has been its connection 
to the hydrodynamic limit and the ability to connect, at least theoretically,  with fluids as diverse as 
the quark-gluon plasma and superfluids \cite{Son:2007vk}. This magnificent feast was accompanied by a 
detailed understanding of transport and linear response as implemented in the context  of the AdS/CFT 
correspondence \cite{Rangamani:2009xk,Hartnoll:2009sz,Herzog:2009xv}.

As is the case in many theoretical frameworks the great success of hydrodynamic descriptions, that is, a description for  large distances and  time-scales carries in itself the seeds of its own limitations. Being an expansion in gradients prevents it from being readily applicable to more extreme regimes. One advantage of the AdS/CFT approach is that on the gravity side there is  {\it a priori} no  limitation on the strength of the nonlinearities.

In this manuscript we want to study the velocity distribution of an external particle in holographic fluids. The main intuition and techniques for the holographic construction of an external particle being dragged in a medium were developed in the context of a particle being dragged in a quark-gluon plasma  which is dual to a classical string embedded in the corresponding supergravity background \cite{Herzog:2006gh,Gubser:2006bz}. We naturally start by revisiting this scenario which has been applied widely to understand the energy loss by quarks in media \cite{Chernicoff:2008sa,Caceres:2006as}, including in confining backgrounds \cite{Mahato:2007zm}.

One of our main motivations is to explore to what extent holographic fluids display properties similar to those of quantum fluids. Take for example quantum turbulence. The term quantum turbulence generically denotes the turbulent motion of quantum fluids. In this class of fluids one finds, among others,  superfluid Helium and atomic Bose-Einstein condensates. These two fluids are characterized by quantized vorticity, superfluidity and two-fluid behavior. Although we do not attempt to construct precise holographic duals of these fluids we turn to the natural question of considering what properties of these quantum fluids are displayed by known holographic fluids.

To orient our inquiry toward a more quantitative goal we consider the experimental 
result \cite{PhysRevLett.101.154501}, which focuses in the velocity statistics as a way to
distinguish between quantum and classical turbulence.  More precisely, by analyzing trajectories 
of solid hydrogen tracers, the authors found that the distributions of velocity in decaying quantum 
turbulence in superfluid ${}^4$He are strongly non-Gaussian with $1/v^3$ power-law tails. These features 
differ from the near-Gaussian statistics of homogeneous and isotropic turbulence of classical 
fluids \cite{PhysRevLett.101.154501}.

One of the longer-term motivations to study quantum turbulence is as follows. There are important conceptual
problems in the traditional approach to quantum turbulence. Namely, unlike classical turbulence which is
studied on the basis of the Navier-Stokes equation, there is no single equation governing the motion of
quantum turbulence. It is more appropriate to say that there is a hierarchy of models at different length
scales, each with its own limitations. The overwhelming microscopic  favorite is the Gross-Pitaevskii 
equation (also called nonlinear
Schr\"odinger) but it misses many effects and requires modifications such as coupling to a
bath \cite{Barenghi25032014}. In this context it is fair to ask: Can gravity provide a relevant
model? As a more long-term goal we expect that a holographic approach might shed some light on the
nature of the equations for quantum turbulence.

There is, indeed, some precedent of exploring holographic models  with the aim of confronting with various
general aspects of fluids. For example,  the authors of \cite{Gubser:2009qf} considered an external heavy 
probe in the zero temperature limit of a holographic superconductor.  Other explorations of turbulence
include \cite{Adams:2013vsa,Green:2013zba,Adams:2012pj,Ewerz:2014tua} and  \cite{Du:2014lwa} which address some general aspects 
of quantum turbulence; a more recent discussion was presented in \cite{Lan:2016cgl}. In this manuscript
we pursue a more modest goal by merely studying, with velocity statistics as the main measure, the holographic
dual of a heavy external particle being dragged in a fluid.

The manuscript is organized as follows. In section \ref{sec:DragGeneral} we briefly review the 
holographic dual of a heavy external particle being dragged in a fluid. We make emphasis on the 
velocity statistics and focus on the properties of the background that might lead to a different  
behavior from the isotropic quark gluon plasma.  In section \ref{Sec:MagPlasma} we consider
a magnetized-quark gluon 
plasma\footnote{Our results here partially overlap with those of~\cite{Finazzo:2016mhm}
which appeared when this work was being concluded.} 
and in section \ref{Sec:SFFlow} we consider a superfluid flow. We conclude in section \ref{Sec:Conclusions}. In  appendix  \ref{App:MagBack} we describe the construction of the magnetic background and some other aspect of the drag while in appendix \ref{App:SFF} we present technical details involved in the construction of the supergravity background dual to the top-down  superfluid flow.

\section{External particle on a quantum fluid holographically: The reasonable persistence of quenching}
\label{sec:DragGeneral}
In this section we study, in very general terms, the behavior of a heavy particle being dragged in a
strongly coupled fluid. We use the holographic dual description which corresponds to a classical string
being dragged in the dual supergravity background.  It is worth pointing out that using a fundamental string, strictly speaking, correspond to a particular type of external heavy particle.  To illustrate the reach of this restriction recall that in  the case of the prototypical duality between strings in $AdS_5\times S^5$ and ${\cal N}=4$ SYM whereby the  classical example of half BPS Wilson loops corresponds, indeed, to a fundamental string with $AdS_2$ wolrdsheet. However, to consider the Wilson loop in higher dimensional representations of $SU(N)$ one needs to consider,  holographically, D3 branes wrapping $AdS_2\times S^2$ and D5 branes wrapping $AdS_2\times S^4$ \cite{Drukker:2005kx,Gomis:2006sb}.  In the case of the correspondence involving confining theories one can naturally consider $k$-string configurations which are composed on the field theory side by $k$ quarks and $k$ anti-quarks; holographically they correspond also to D-branes \cite{Herzog:2001fq}. Thus, our analysis below is limited to external heavy particles of a very particular  type.

Our starting point is a five-dimensional subspace of a supergravity background; this should be viewed
as focusing in the four field theory dimensions of the  dual to a field theory plus the most needed
holographic direction. 
\be
\label{metric}
ds^2=-G_{00}\,dt^2+G_{xx}\,dX^2+2G_{0x}\,dt dX+G_{rr}\,dr^2+dX_2^2+dX_3^2,
\ee
where the
metric components $G_{MN}$ are functions of the radial
coordinate $r$ only. Following \cite{Herzog:2006gh,Gubser:2006bz}, we assume the worldsheet to be
embedded as $t=X_0$
and $\s=r$ and we allow for
\be\label{encaje}
X=vt+\xi(r),
\ee
which means that the end of the string is moving with velocity $v$. In the field theory side we
interpret the end of the
string as an external heavy particle moving with velocity $v$; the canonical example 
of \cite{Herzog:2006gh,Gubser:2006bz} corresponding to a heavy external quark moving in a 
strongly coupled quark-gluon plasma.  The Nambu-Goto action describing this embedding is
\be
\label{Eq:Action}
S=-\frac{1}{2\pi\alpha'}\int dt\, d\sigma\, \sqrt{G_{00}\,G_{rr}
-G_{xx}\,G_{rr}\,v^2-2G_{0x}\,G_{rr}\,v+(G_{00}\,G_{xx}+G_{0x}^2)\,\xi '^2}\,.
\ee
Since the action does not depend explicitly on $\sigma$, one has that the
conjugate momenta $\frac{\partial {\cal L}}{\partial \xi '}$ is a constant,
where ${\cal L}$ is the Lagrangian density, so
\be
\Pi _{\xi}= \frac{\partial {\cal L}}{\partial \xi '}=
-\f{(G_{00}\,G_{xx}+G_{0x}^2)\,\xi'}{\sqrt{G_{00}\,G_{rr}
-G_{xx}\,G_{rr}\,v^2-2G_{0x}\,G_{rr}\,v+(G_{00}\,G_{xx}-G_{0x}^2)\,\xi'^2}},
\ee
which can be rearranged to obtain
\be\label{xieq}
\xi '=\Pi_{\xi}\,\sqrt{\f{G_{rr}}{G_{00}\,G_{xx}+G^2_{0x}}}\sqrt{\frac{G_{00}
-G_{xx}\,v^2-2G_{0x}\,v}{G_{00}\,G_{xx}+G_{0x}^2-\Pi _{\xi}^2}}.
\ee
A key simplifying observation is that, since $\xi '$ cannot be imaginary, we need both expressions in the 
numerator and the denominator to flip signs simultaneously. This fixes $\Pi _{\xi}$ to be
\be
\label{Pi}
\Pi_{\xi}^2=G_{00}(r_*)\,G_{xx}(r_*)+G_{0x}(r_*)^2.
\ee
Here, $r_*$ denotes the radial coordinate satisfying
\be
\label{u*}
G_{xx}(r_*)v^2+2G_{0x}v-G_{00}(r_*)=0.
\ee
The rate of change of momentum is calculated
to be
\be
\f{dp_1}{dt}=\sqrt{-g}\,T^{r}{}_{x}\,,
\ee
where
\be
T^r{}_{x}=-\frac{1}{2\pi\alpha '}\,G_{x\nu}\,g^{r\alpha }\,\partial_{\alpha}X^{\nu}\,.
\ee
Here, $G_{ij}$ denotes the metric in (\ref{metric}), while
$g_{ij}$ denotes the induced metric on the world-sheet. After some
algebraic simplifications, we obtain
\be
\label{DF1}
\frac{dp}{dt}=-\frac{1}{2\pi\alpha '}\Pi_{\xi}\,.
\ee
Thus, $\Pi_{\xi}$ determines the rate of momentum loss.

\subsection{Velocity statistics}

To connect with  experimental studies let us consider one relevant quantity which plays a central role in distinguishing the different types 
of turbulences.  
An important, measurable quantity in turbulent flows is the probability of observing a velocity 
between $v$ and $v+dv$, denoted by $P_v(v)dv$. Let us also denote the uniform probability of taking a 
measurement at time between $t$ and $t+dt$, $P_t(t)dt$.
Assuming
    \be
    P_v(v)=P_t\left(t(v)\right)\left|\frac{dt}{dv}\right|,
    \ee    
one arrives at
\be
    P_v(v)\propto \left|\frac{dt}{dv}\right|.
\ee
The most natural expectations is that this statistics follows a Gaussian distribution. The velocity statistics is central in experimental studies of fluid flows. It is particularly useful as a potential way of characterizing quantum turbulent flows as discussed in \cite{PhysRevLett.101.154501} where it was  discovered that for superfluid ${}^4$He it  deviates from the near-Gaussian statistics of homogeneous isotropic turbulent classical flows. 

Most of the experimental analysis using velocity statistics is limited to non-relativistic regimes. We will naturally  extend the above definition of velocity statistics to the relativistic  regime as by writing momentum rather than velocity:
\be
P_v(v)\propto m \frac{dt}{dp}.
\ee
At this point we simply observe that such appropriately generalized velocity statistics is mathematically equivalent to the expression for drag force. Namely, using Eq. (\ref{DF1}) , one obtains

\be
P_v(v)\propto -2\pi \alpha'\,m\,\frac{1}{\Pi_\xi}.
\ee
The drag force can be computed directly using holography and this is the relation we exploit throughout the manuscript. We now address  this question using holography since we have verified that the appropriately generalized velocity statistics is related to the drag force.

\subsection{The dragged string  in generic backgrounds}

With the aim of developing our intuition about the interplay between  the geometry and 
the velocity statistics here we revisit a few examples, some of them fairly canonical.

\subsubsection{The dragged string  in the quark-gluon plasma}

One of the  most interesting consequences of studying the string moving in a holographic quark-gluon
plasma is its explanation of the jet-quenching phenomena, which is studied using the methods that have been described so far throughout this section.

As in \cite{Herzog:2006gh,Gubser:2006bz}, we shall consider a string moving in the black D3-brane
geometry dual to ${\cal N}=4$ SYM at finite temperature, whose 
near horizon geometry can be approximated by
\be
\label{Eq:FiniteT_horizon}
G_{00}\approx  g_0 \,(r-r_h), \qquad G_{rr}\approx f_0\,(r-r_h)^{-1}, 
\qquad G_{xx}\approx g_x\, r^2\,,
\ee
For this metric Eq. (\ref{u*}) results in
\bea
v^2&=& \frac{g_0(r_*-r_h)}{g_x\, r_*^2}.
\eea
We can solve for $r_*$ to obtain
\be
r_*=\frac{g_0}{2g_x \, v^2}\left(1\pm \sqrt{1-4\frac{g_x}{g_0}\,r_h\,v^2}\right)\,,
\ee
which in the small velocity limit becomes
\be
r_*\approx r_h+\frac{g_x}{g_0}\,r_h^2\, v^2,\label{smallv}
\ee
leading to
\be
\label{Eq:PI_universal}
\Pi _{\xi}\approx g_x\,r_h^2\,v\left(1+2\frac{g_x}{g_0}r_h\, v^2\right).
\ee
The leading term in this regime is always linear and we conclude that
\bea
\frac{dv}{dt}&=&-\frac{g_x\,r_h^2}{2\pi \alpha'}\,v\left(1+2\frac{g_x}{g_0}r_h\, v^2\right)\,,
\eea
and keeping the 
leading linear term only we find that 
\be
v(t)=v_0\, e^{-kt}\,,
\ee
which is the typical damping behavior (jet quenching). 

We should compare this result with the experimentally observed velocity statistics in
quantum turbulence \cite{PhysRevLett.101.154501}, we obtain  
\be
\frac{dv}{dt}=-k\,v^3, \longrightarrow v(t)=\frac{1}{\sqrt{2k}\sqrt{t-t_0}}.
\ee

It is worth trying to understand the universality of the behavior 
in Eq. (\ref{Eq:PI_universal}) from the holographic point of view.
There are two main inputs that enter in the result. First, the assumption that the metric
factors behave as in Eq. (\ref{Eq:FiniteT_horizon}), which amounts to having a supergravity
background describing a finite temperature solution. The second assumption, which is milder,
is that the value $r_*$ is close to the horizon. 

\subsubsection{Quantum Critical Point}

We will show that the above linear behavior is fairly generic. 
Indeed, one obtains a sort of no-go theorem if one stays with 
``typical finite temperature  horizons.''  
We will discuss the following example in more detail in the coming sections and will explore it 
numerically but for now let us focus on its near horizon behavior and show that, in this limit, the velocity statistics
do not change substantially.  The background in question was advanced
by \cite{D'Hoker:2009mm} as a model of a potential quantum critical point. We find the near horizon 
expansion to be
\bea
G_{00}&=&(12-4b^2)\,(r-1)^2, \nonumber \\
G_{0x}&=& 2b\,(r-1)\,, \qquad G_{xx}=1\,.
\eea
Therefore
\be
v=(b\pm \sqrt{3})\,(r-1)\,,
\longrightarrow
\Pi_\xi =2\sqrt{3}(r-1)\sim v\,,
\ee
implying the persistent quenching discussed before. 

\subsubsection{Superfluids}

Another interesting system is a holographic superfluid flow as that constructed
in~\cite{Arean:2010wu}.
We will discuss this model numerically in more detail in section 
\ref{Sec:SFFlow}; here we focus on the persistence of ``quenching.''
It was found in~\cite{Arean:2010wu} that the following metric is general enough
to accommodate a holographic superflow: a holographic superfluid with a nonzero
superfluid velocity (along $x$ in this case).
%
%
\be
ds^2 =-\frac{r^2 f}{L^2}\,dt^2 +\frac{L^2 h^2}{r^2 f}\,dr^2 
-2C\,\frac{r^2}{L^2}\,dt\,dx +\frac{r^2\,B}{L^2}\,dx^2 +\ldots\,.
\ee
The Lagrangian for the corresponding string motion is precisely of the general form 
Eq.~(\ref{Eq:Action}) times an overall factor
$Q\equiv Q(\psi)=\cosh(\psi/2)$
resulting from the uplift to 10d IIB String Theory~\cite{Gubser:2009qm,Gubser:2009qf}
of the five-dimensional model \eqref{IIBac}
where $\psi$ is the real scalar field dual to the order parameter of the superfluid
(see section~\ref{Sec:SFFlow} and appendix~\ref{App:SFF}).
The analog of the classical string solution is 
\be
\xi'=\frac{h\, L^4}{r^4\, (fB+C^2)}\,\Pi_\xi\,\sqrt{\frac{1-\frac{B}{f}\,v^2+2\frac{C}{f}\,v}
{Q^2-\frac{L^4}{r^4\,(f\,B+C^2)}\,\Pi_\xi^2}}\,.
\ee
Given the near horizon geometry of this case,
\bea
f&=&f_1^H(r-r_H)\,, \nonumber \\
h&=&h_0^H, \qquad B=B_0^H\,,\nonumber \\
C&=&C_1^H(r-r_H)\,, 
\label{nearHfiniteT}
\eea
we see that arguments similar to those leading to Eq. (\ref{smallv}) also apply, so we can conclude again that small velocities are ``concentrated'' near the horizon, and
\be
\Pi_\xi \approx Q_*\,\frac{r_*^2}{L^2}\,B_0\, v\,,
\ee
where $Q_*=\cosh(\psi/2)|_{r=r_*}$.

The key observation is that whenever we use the horizon asymptotics corresponding to finite 
temperature, and hence have a near horizon metric 
of the form
(\ref{nearHfiniteT}), we will obtain exponential decaying velocities.

Let us mention here that the factor $Q$ in front of the action of a string in the
type IIB superfluids~\cite{Gubser:2009qm,Arean:2010wu} slightly modifies the expression
\eqref{Pi} for the drag. Indeed, in those systems, which will be studied in 
section~\ref{ssec:SFFlow}, one has
\be
\Pi _{\xi}^2=Q_*\left(G_{00}(r_*)\,G_{xx}(r_*)+G_{0x}(r_*)^2\right)\,.
\label{eq:dragsf}
\ee

\subsubsection{Lifshitz geometries}
\label{sec:lif}
Among the class of holographic liquids it is interesting to consider Lifshitz models. 
Assuming the classical string explores a deep IR geometry which can be modeled as a Lifshitz
geometry we have 
\be
ds^2 = L^2 \left(-r^{2z}\,dt^2 +r^2\, d\vec{x}^2 +\frac{dr^2}{r^2}\right)\,.
\ee
In this case $G_{00}= r^{2z}, \quad G_{xx}=r^2$, which leads to 
\be
\Pi_\xi=v^{\frac{z+1}{z-1}}.
\ee
In the case of $z=2$ we obtain
\be
\frac{dv}{dt}=-kv^3, \qquad P_{v} \sim |v|^{-3}\,.
\ee

\subsubsection{Engineering geometries dual to quantum turbulence}

We shall now apply 
a reverse engineering approach, namely to look for  the behavior of 
the metric functions near the turning point of the classical string that leads to the kind of
velocity
statistics compatible with quantum turbulence.
A generical velocity distribution is given by
\be
P_{v}\sim \frac{1}{\Pi_\xi}.
\ee
Assuming a metric of the form: $G_{00}=r_*{}^{\Delta_1},\; G_{xx}=r_*{}^{\Delta_2},\; G_{0x}=0\,$, 
we find
\be
\Pi_\xi=v^{\frac{\Delta_1+\Delta_2}{\Delta_1-\Delta_2}},
\longrightarrow P_{v}\sim |v|^{\frac{\Delta_1+\Delta_2}{\Delta_2-\Delta_1}}.
\ee
Getting $P_v\sim |v|^{-3}$ requires $\Delta_1=2\Delta_2$.

\section{Dragging a quark in a magnetized quark-gluon plasma}\label{Sec:MagPlasma}

Our main motivation, as stated in the introduction,  is to learn about the properties of 
holographic quantum fluids. There are, however,  important reasons to pause for a detailed
discussion of a magnetized quark-gluon plasma. Understanding the physics of the quark-gluon 
plasma formed in highly energetic particle collisions, like those at RHIC and the LHC experiments, 
is still a challenge. There are a large number of experimental outcomes that still  lack a satisfactory
explanation or are simply beyond the reach of standard  field theory techniques. One of the main reasons 
resides, to most people's surprise, in the fact that this plasma remains a strongly coupled system 
\cite{Shuryak:2003xe,Shuryak:2004cy}, making  perturbative methods inapplicable.  
Given the scarcity of non-perturbative methods and the inherent Euclidean nature of
the lattice calculations, dynamical processes typically escape the kind of analysis
that can be performed by these approaches, providing a strong motivation to the 
community that works in the gauge/gravity correspondence to approach this topics using holographic methods. 
A review of many of the physical processes that have been studied using this correspondence can be
found in \cite{CasalderreySolana:2011us}.

Among the physical phenomena that have been analyzed by holographic methods, the drag force
experienced by a quark traveling across the plasma has received a lot of attention,
starting with \cite{Herzog:2006gh} and \cite{Gubser:2006bz} and followed by a large
number of generalizations.

Another important motivation to revisit the computation of the drag force in the presence of 
a strong magnetic field, is because for non-central collisions such a field is expected to be 
present in the resulting plasma, and it was suggested \cite{Basar:2012bp} that this field could 
be responsible for the enhancement of direct photon production in heavy ion collisions 
reported in \cite{Wilde:2012wc} and the anisotropy of their distribution reported in \cite{Adare:2011zr}. 
Not too long ago \cite{Arciniega:2013dqa} one of us was  able to qualitatively reproduce these peculiarities 
of the photon spectrum performing gravitational calculations on the background originally suggested in
\cite{D'Hoker:2009mm} which places the plasma in the presence of a strong constant magnetic field.

The presence of the magnetic field singles out a particular direction, so it clearly breaks the isotropy
in the gauge theory. In the past the drag force has been studied for anisotropic plasmas sourcing the
anisotropy in different ways. For example in  \cite{NataAtmaja:2010hd} the rotational symmetry was
broken by the angular momentum of a black hole in the gravitational solution and in
\cite{Chernicoff:2012iq, Giataganas:2012zy} the source of the anisotropy was a position-dependent axion on
the gravity side equivalent to a position-dependent theta term in the dual gauge theory. 
The presence of a magnetic field in the background of the field theory that should be used 
to study the quark-gluon plasma seems to be undeniable, so we would like to see the effects 
of it also as a source for anisotropy. In particular in \cite{Chernicoff:2012iq} the authors 
analyzed the generality of their result with respect to the physics behind the anisotropy, 
and speculated  that given some reasonable assumptions, their  results should be reproduced 
within any implementation of anisotropy independently of its explicit form. In \cite{Giataganas:2012zy} the author analyzes the effect of the anisotropy over some observables other than the drag force and includes a discussion about different anisotropic models. Recently, a study of a heavy impurity moving longitudinal to the direction of an external magnetic
 field in an anomalous chiral medium  was also considered \cite{Sadofyev:2015tmb}.

The construction that we will use is identical to the one  carried out in \cite{Arciniega:2013dqa} 
to get a gravitational dual to a field theory with a constant magnetic field. We provide more details 
of the construction of the solution in appendix  \ref{App:MagBack}, where we describe a way to carry
this construction to make it very easy to work with the desired background. The starting point and 
general idea is to take the gravitational construction performed  in \cite{D'Hoker:2009mm} and
numerically compute it in such a way that we get a one parameter family of gravitational
configurations that interpolates between two analytic solutions of interest to us.

The background we study is given by a metric that asymptotes to  $AdS_5$  and that has a 
magnetic field tangent to the boundary directions of this limiting geometry. To this end we 
seek solutions to the Einstein-Maxwell equations that fit the Ansatz of \cite{D'Hoker:2009mm},
provided by a metric of the form
\begin{equation}
ds^2=-U(r)dt^2+V(r)(dx^2+dy^2)+W(r)dz^2+\frac{dr^2}{U(r)},\label{metans}
\end{equation}
and a field strength given by
\begin{equation}
F=B\, dx \wedge dy.\label{Fans}
\end{equation}

Solutions of this type preserve the invariance under spacetime translations and under rotations
in the $(x,y)$ plane. The magnetic field (\ref{Fans}) in any metric of the form (\ref{metans})
satisfies Maxwell equations identically, while the gravitational equations  can be followed 
in appendix \ref{App:MagBack}.

As we explain in detail in appendix \ref{App:MagBack}, we provide our results in terms of 
quantities that are meaningful on the field theory side. For example, for the value of the 
magnetic field we must consider $b\equiv B/V(r\rightarrow\infty)$ to take in to account the 
area norm of the 2-form field strength, moreover, for the magnitudes to be relevant we also 
focus on dimensionless quantities such as $b/T^2$, where $T$ is the temperature of the black
hole and the dual field theory.

The dynamics of a quark moving through a quark-gluon plasma has been studied extensively
following the pioneering works  \cite{Gubser:2006bz,Herzog:2006gh}. However, the existence 
of a magnetic field in the dual field theory has received little attention. The experimental
evidence gathered on accelerators such as RHIC and LHC together with the recent 
work \cite{Arciniega:2013dqa} (and more recently \cite{Finazzo:2016mhm}) highlight
the relevance of taking into account this magnetic field. 

It is worth mentioning that what we report in sections \ref{para} to \ref{perpen} is the effect, 
on the drag force, of the reaction of the quark-gluon plasma to the presence of a magnetic field using a probe particle that is not directly coupled to $F$.
That there is not such a direct coupling can be seen from the embedding in 10 dimensional type IIB supergravity 
\cite{Patino2016} of the background we are using. In this uplift the vector potential $A$, of which $F$ is the field strength, is encoded simultaneously in three $U(1)$ symmetries of an internal 5-sphere, and as a deformation of the IIB five form $F_5$. The $F_5$ does not couple directly to the two dimensional worldsheet of the string which tip we use to model the probe particle. As far as the 10 dimensional metric is concerned, the shift associated with $A$ in three of the compact directions is a volume preserving transformation, that once the appropriate tetrad is introduced, keeps the directions of the internal space perpendicular to those of the extended five dimensional space that we use in our calculations. The use of the aforementioned tetrad makes apparent that  the dynamics in the five dimensional extended space is factored out from any wrapping of the worldsheet along the $U(1)$ directions. In the particular case of motion perpendicular to the magnetic field, such wrapping does not even exists.

\subsection{A quark moving parallel to the magnetic field}\label{para}

To include the effects of the magnetic field on the quark's dynamics we consider the analysis of
the probe quark presented in the previous section. The supergravity metric, (\ref{metans}), 
incorporates the back-reaction due to magnetic field (\ref{Fans}). This is all that is needed, 
since, as explained before, the embedding of this solution in 10 dimensional IIB 
supergravity \cite{Patino2016} shows that the string does not couple to $F$ directly, 
and furthermore, the solution of the string moving in this lift can be obtained by solving
the equations of motion in our five dimensional background. The quark moves with a constant
velocity, exactly as the Ansatz (\ref{encaje}).  The differential equation which we have to 
solve results from Eq. (\ref{xieq}) into 
\begin{equation}\label{diffxi}
\xi'(r)=\pm\frac{2\pi\alpha'\Pi_z}{\sqrt{U W}}\sqrt{\frac{1-\frac{W}{U}v^2}{UW-{2\pi\alpha'\Pi_{z}}^2}}.
\end{equation}

\begin{figure}[htb]
\centering
\includegraphics[width=.8\linewidth]{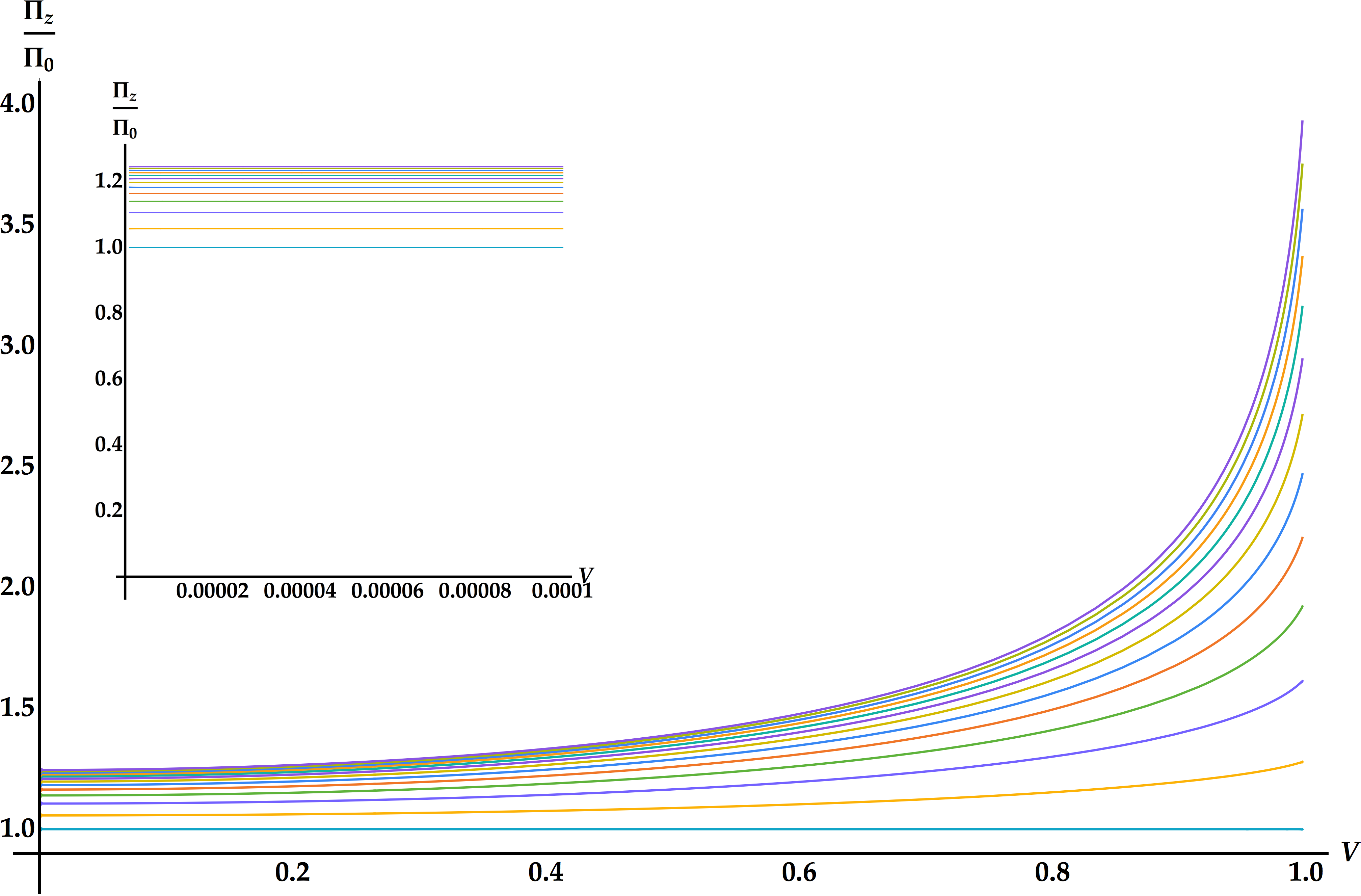}   
\caption{$\Pi_z$ as a function of the velocity, normalized with respect to the 
analytic result $\Pi_0(v)\equiv\Pi_z(v,b=0)$ for 
$b/T^2=0, 14.74, 30.31, 47.88, 67.14, 87.87, 109.89, 133.06, 157.27, 182.44, 208.50, 235.39, 263.06$ 
with larger values of $b/T^2$ corresponding with higher lines. Notice that the lowest line demonstrates 
the consistency of our numerical results with previous analytic findings. The inset shows the behavior
of the same plots for velocities very small compared to 1.}\label{PznormV}
\end{figure}
\begin{figure}[htb]
\centering
\includegraphics[width=.8\linewidth]{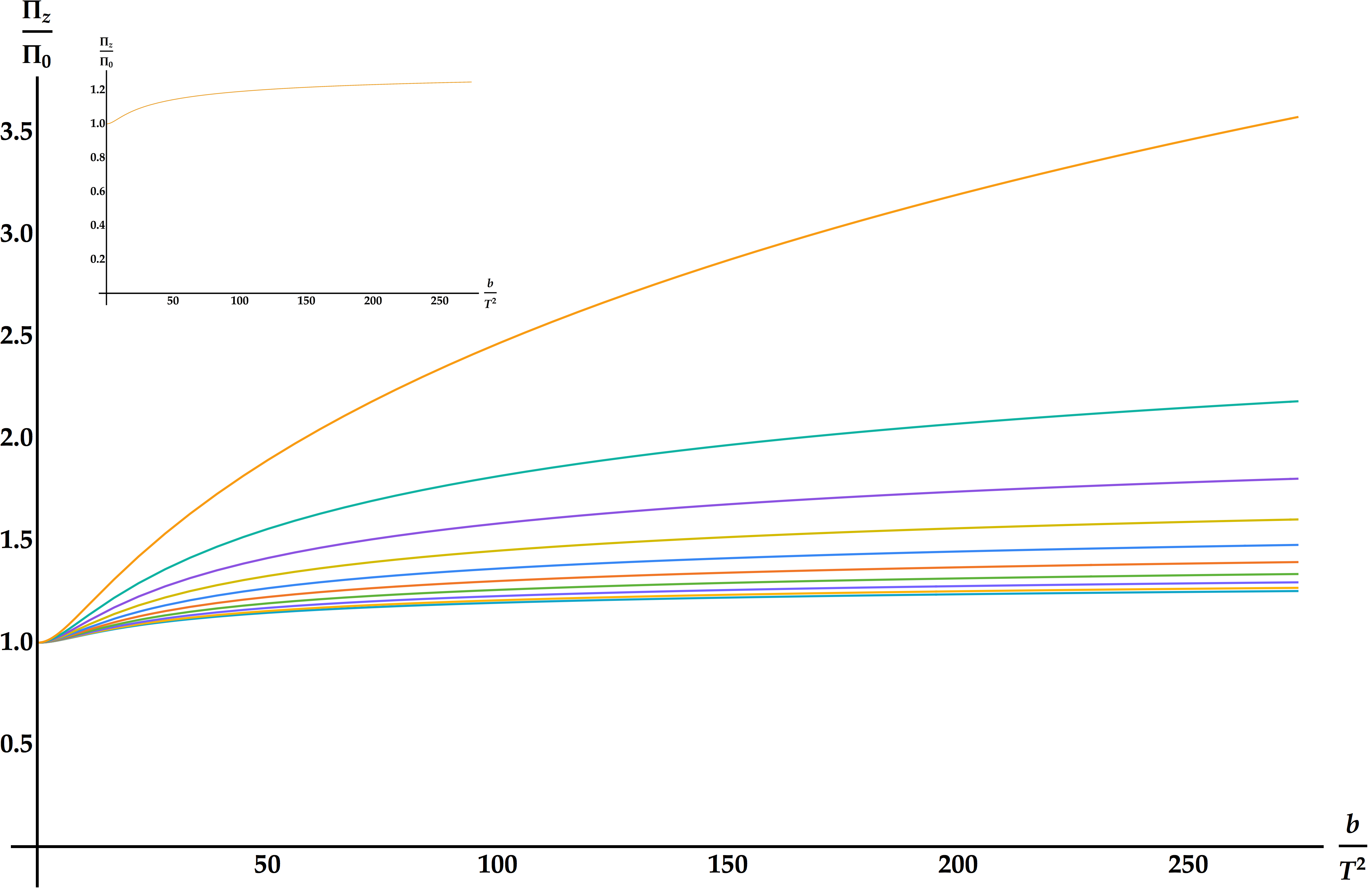} 
\caption{$\Pi_z$ as function of $b/T^2$, 
normalized with respect to the analytic result $\Pi_0(v)\equiv\Pi_z(v,b=0)$ 
for velocities 27/250, 103/500, 38/125, 201/500, 1/2, 299/500, 87/125, 397/500, 223/250, 99/100. 
The inset shows the plots for velocities 
1/100000, 1/50000, 3/100000, 1/25000, 1/20000, 3/50000, 7/100000, 1/12500, 9/100000, 1/10000. 
Larger values of the velocity correspond with higher lines. Notice that all the plots coinciding at value 
1 for $b/T^2=0$ demonstrate the consistency of our numerical results with previous analytic findings.}
\label{PznormB}
\end{figure}

The relevant information about the quark's dynamics is contained in the functional dependence of the 
integration constant $\Pi_z(v,b)$, see Eq. (\ref{DF1}). This behavior has been plotted in figures
\ref{PznormV} and  \ref{PznormB}, where $\Pi_z(v,b)$ is presented as a function of the velocity 
for a number of intensities of the magnetic field, and also as a function of $b/T^2$ for a number
of velocities. We provide plots for the normalized value of $\Pi_z(v,b)$ with respect to the
analytic result for $\Pi_0(v)\equiv\Pi_z(v,b=0)$ \cite{Gubser:2006bz,Herzog:2006gh},
\begin{equation}\label{czgub}
\Pi_0(v)= \frac{{r_h}^2}{L^2} \frac{v}{\sqrt{1-v^2}},
\end{equation}
where $r_h$ is the position of the horizon and $L$ the AdS radius of an AdS-Schwarzschild background.

In the normalized plot \ref{PznormV} we can see that the lower line, given by our numerical result
for $b=0$, is a constant function at unit value, providing a benchmark that connects our numerical
solutions to the previous solution \cite{Gubser:2006bz,Herzog:2006gh}, since certainly, in the 
absence of magnetic field we should match (\ref{czgub}) exactly,  as the horizontal line in 
Fig.~\ref{PznormV} demonstrates. In appendix \ref{App:MagBack} we also plot the values of $\Pi_z(v,b)$ 
without normalization.

Let us describe the behavior shown in the plots in more detail. We see that in all cases the reaction 
of the plasma to the presence of a magnetic field increases the drag force experienced by a quark traveling through it.
We also see that the magnetic field causes the drag force to have a higher than linear dependence on the
velocity  in the relativistic region.  From the inset in Fig.  \ref{PznormV} we verify that, nonetheless,
for non-relativistic velocities  the drag continues to be linear in $v$. This is, again, the persistent 
behavior described analytically in the previous section. What we conclude is that   the constant of
proportionality increases with the value of $b/T^2$.

Regarding the dependence of the drag on the magnetic field, that is, with  respect to $b/T^2$, it 
is worth noticing that the plots are shown to be horizontal at $b/T^2=0$, that is, 
$[\partial_{b/T^2}\Pi_z]_{b/T^2=0}=0 \:\:  \forall \: v$.

Some other properties of the response to the presence of the magnetic field will be discussed in 
section \ref{diffzxy}, where we will be concerned with those properties that, as the next section
will show, depend on the direction of the movement of the quark.

\subsection{A quark moving perpendicular to the magnetic field}\label{perpen}
We now study a quark moving perpendicularly to the magnetic field. The analysis is the same as in the
previous section, and all the remarks, including the ones about the 10 dimensional embedding, still apply. 
The differential equation describing the quark motion is
\begin{equation}\label{diffxixy}
\xi'(r)=\pm\frac{2\pi\alpha'\Pi_i}{\sqrt{U V}}\sqrt{\frac{1-\frac{V}{U}v^2}{UV-{2\pi\alpha'\Pi_i}^2}},
\end{equation}
where now $\Pi_i$ stands for either $\Pi_x$ or $\Pi_y$, and we notice that the only difference with 
Eq. (\ref{diffxi}) is the exchange of $W$ for $V$.

\begin{figure}
  \centering
  \includegraphics[width=.8\linewidth]{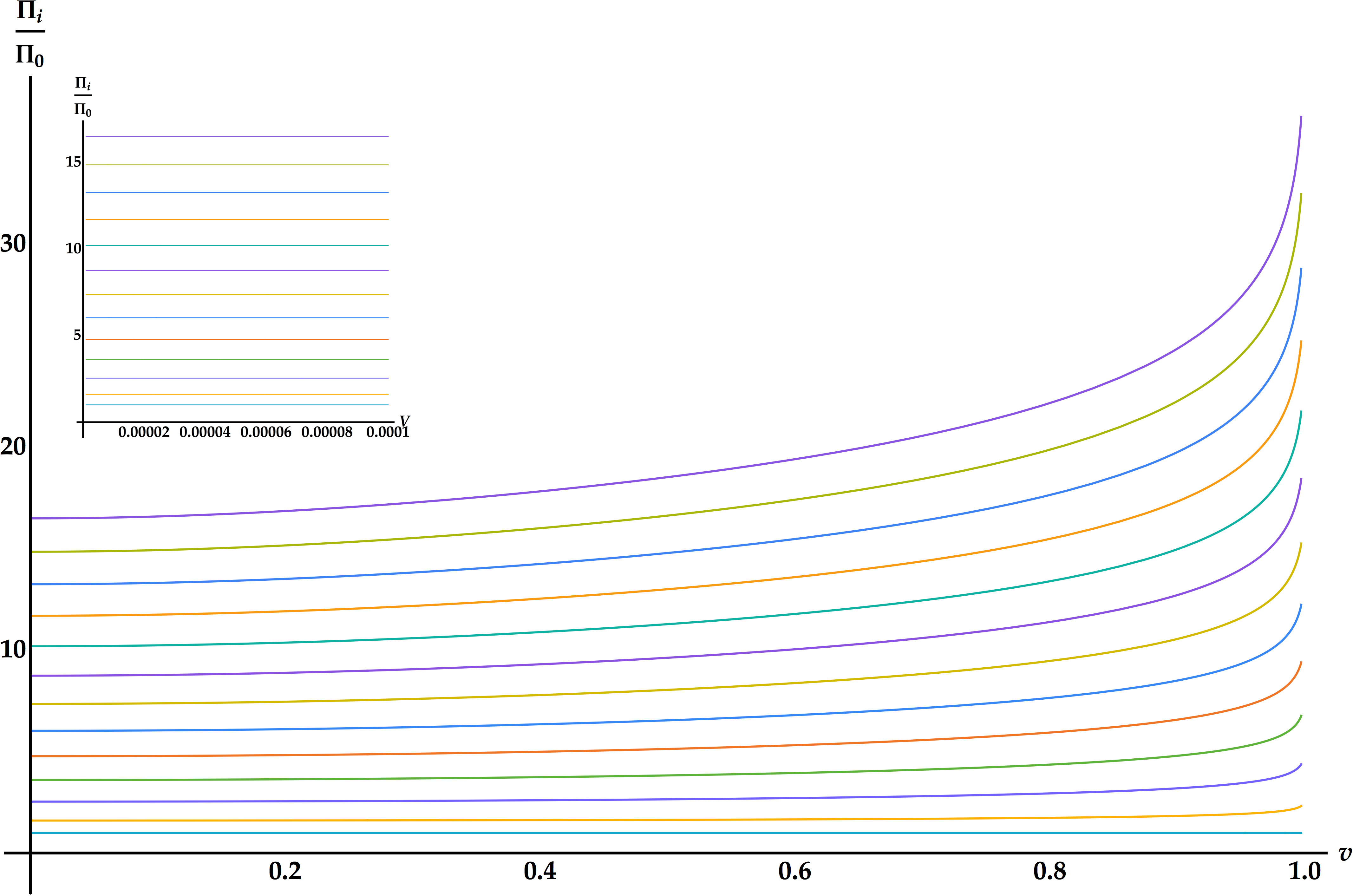}

\caption{$\Pi_{i=x,y}$ as function of the velocity, normalized with respect to the analytic result $\Pi_0(v)\equiv\Pi_z(v,b=0)$ for $b/T^2=0, 14.74, 30.31, 47.88, 67.14, 87.87, 109.89, 133.06, 157.27, 182.44, 208.50, 235.39, 263.06$ with larger values of $b/T^2$ corresponding with higher lines. Notice that the lowest line demonstrates the consistency of our numerical results with previous analytic findings. The inset shows the behavior of the same plots for velocities very small compared to 1.}
\label{PxynormV}
\end{figure}

\begin{figure}[htb]
\begin{center}
\includegraphics[width=.8\linewidth]{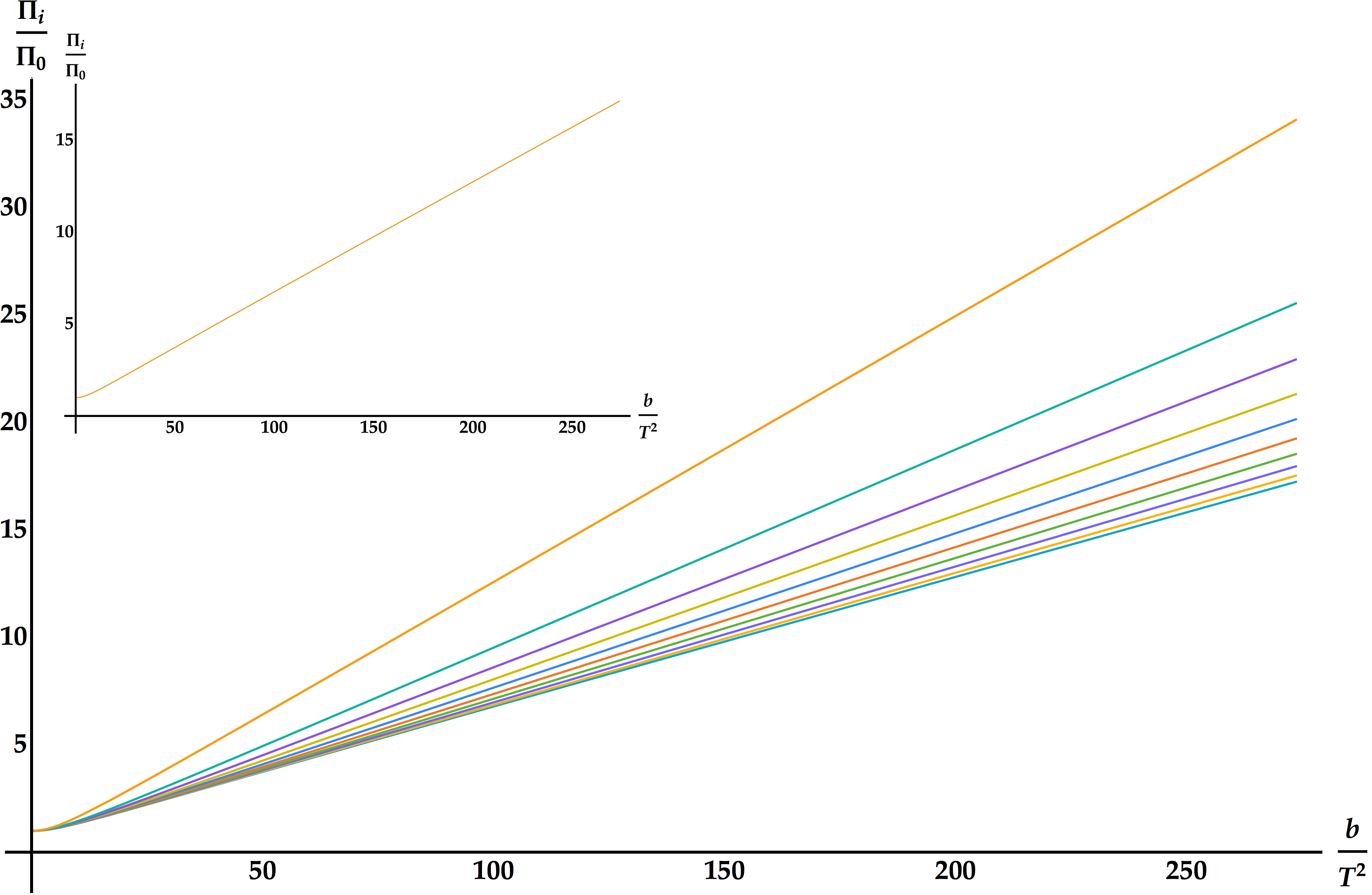} 
\captionof{figure}{$\Pi_{i=x,y}$ as function of $b/T^2$, normalized with respect to the analytic
result $\Pi_0(v)\equiv\Pi_z(v,b=0)$ for 
velocities 27/250, 103/500, 38/125, 201/500, 1/2, 299/500, 87/125, 397/500, 223/250, 99/100. 
The inset shows the plots for velocities 
1/100000, 1/50000, 3/100000, 1/25000, 1/20000, 3/50000, 7/100000, 1/12500, 9/100000, 1/10000. 
Larger values of the velocity correspond with higher lines. Notice that all the pots coinciding a 
value 1 at $b/T^2=0$ demonstrate the consistency of our numerical results with previous analytic findings.}
\label{PxynormB}
\end{center}
\end{figure}

We present the results in the same fashion as in the previous subsection. In Fig. \ref{PxynormV} 
we present the normalized value of the drag as a function of the velocity and in Fig.  \ref{PxynormB}
we present the normalized value of the drag as a function of the magnetic field.  It is  an important
numerical check that we find the same agreement with the analytic results for $b=0$.

The general behavior of the drag force in the direction perpendicular to the magnetic field can be
extracted from the information condensed in the plots in Fig.  \ref{PxynormV}  and Fig. \ref{PxynormB}. 
For completeness we also plot the non-normalized values in appendix \ref{App:MagBack}. The characteristics
that coincide with the case of movement in the parallel direction have already been described in 
subsection \ref{para}, while most importantly the differences will be discussed in the 
following subsection.

\subsection{Differences between movement parallel and perpendicular to $B$}
\label{diffzxy}

There are at least two key differences between movement of the quark parallel and perpendicular
to the direction of the magnetic field. The first is the asymptotic behavior of the drag force
as $b/T^2$ increases, that proves to get to a maximum for the parallel movement as shown in 
Fig.~\ref{PznormB}  (and also in Fig.~\ref{PzB} in appendix \ref{App:MagBack}) 
meanwhile the force in perpendicular direction grows without bound as seen in Figs.~\ref{PxynormB} 
(and also in Fig. \ref{PxyB} in appendix \ref{App:MagBack}).

The second point is the general magnitude of both forces. The force $\Pi_{i=x,y}$ is always much 
larger than $\Pi_z$, at least two orders of magnitude for a reasonable intensity of the magnetic field. 
This behavior reflects the RG flow between a quantum 3+1 field theory (QFT) and a 1+1 QFT. From a 
semi-classical point of view, one can think that the charged particles move around the lines of 
the magnetic field and the lowest energy modes move along the direction of the magnetic field. 
The magnitude of $\Pi_x$ tells us that it is harder to move in this direction than parallel
to the magnetic field. Evidence of this behavior has been found in a superconducting system in which
the extent of the condensate reduces to a line as the magnetic field grows \cite{Albash:2008eh} and 
by studying the algebra of correlators in  \cite{D'Hoker:2010hr,D'Hoker:2016wgl}. 


In order to have a more quantitative description of how the velocity dependence of $\Pi_z$ changes due
to the magnetic field, we adjust straight lines for non relativistic velocities to logarithmic plots
(not displayed). Assuming that  the dependence of $\Pi_z$ on the velocity can be written as 
$\Pi_z(v,b)=a(b)v^{m(b)}/\sqrt{1-v^2}$, so that the slope of the linear fit  is the exponent $m(b)$
we found that $1.035 \le m(b)\le 1.048$ for a range of $b$ changing over two orders of magnitude.

It is clear that the impact of the magnetic field on the slope is very small. We have also considered 
similar estimates for motion in the perpendicular to the magnetic field direction but with similar 
results. Tracking  the behavior of $m(b)$ for both, parallel and perpendicular motion indicates that 
the increase of $m$ is very small and bounded from above as $b$ gets arbitrarily large.

\section{A top-down holographic superfluid flow }
\label{Sec:SFFlow}

Within the context of the AdS/CFT correspondence the phenomenon of superfluidity as characterized by the 
spontaneous breaking of a global $U(1)$ symmetry at sufficiently low temperatures corresponds 
to the existence of certain hairy black holes in AdS. These models have already reproduced many
important properties, including transport properties  \cite{Hartnoll:2009sz,Herzog:2009xv,Horowitz2011}.

There are many simplified models that capture the central physics of superfluidity holographically. 
For the purpose of studying the motion of an external particle in a holographic superfluid we choose 
to work with a model that admits a UV completions via embedding into string theory as first proposed 
in \cite{Gubser:2009qm}. We follow the notation and treatment developed in \cite{Arean:2010wu} with
some minor improvements in the numerical approach. 

We proceed to present a brief introduction to the background, more details can be found in appendix 
\ref{App:SFF}.  As stated above, the key model is a consistent truncation of type IIB supergravity 
which has the structure of an Einstein-Maxwell (plus Chern-Simons) system in five dimensions coupled to
a charged scalar field with a non-trivial potential. The action reads
\bea
\label{IIBac}
S_{IIB}&=&\int d^5x \sqrt{-g}\bigg[R-\frac{L^2}{3}F_{ab}F^{ab}
+ \frac{1}{ 4}\left(\frac{2L}{ 3}\right)^3 \epsilon^{abcde}F_{ab}F_{cd}A_e
+\nonumber
\\
&&- \frac{1}{2}\left((\partial_a \psi)^2 + \sinh^2 \psi(\partial_a
\theta -2 A_a)^2-\frac{6}{ L^2}\cosh^2\left(\frac{\psi}{2}\right)(5-\cosh \psi)\right)\bigg]\,.  
\eea 
Here,
$\epsilon^{01234}=1/\sqrt{-g}$, and we have written the charged
(complex) scalar by splitting the phase and the modulus in the form
$\psi e^{i\theta}$.
Notice that the leading terms in the scalar potential are
\be
V(\psi)={12\over L^2}-{3\psi^2\over 2 L^2}+\dots\,,
\label{eq:scpot}
\ee
which correspond to the AdS cosmological term and the scalar mass term respectively.
In $d=4$, a scalar with $m^2=-3$
has a non-normalizable leading falloff at the AdS boundary, which corresponds to the source of a dual
field theory operator ${\cal O}$ of dimension $\Delta = 3$.
The solutions dual to a superfluid phase 
are those where the source of this scalar operator vanishes, while
its VEV, given by the subleading fall-off of $\psi$, does not.
They spontaneously break the $U(1)$ gauge
invariance realized by the gauge field $A$, which on the boundary theory corresponds
to a global $U(1)$ symmetry whose spontaneous breaking constitutes a superfluid phase transition.

As usual, to study a setup at finite charge density we must switch on the temporal component
of the gauge field. Additionally, as in \cite{Basu:2008st,Herzog:2008he}, a finite superfluid velocity is turned on via
a non-vanishing spatial component of the gauge field.
Consequently, we consider the following Ansatz for the gauge field and the scalar
\bea A=A_t(r)\, dt + A_x(r) \,dx\;, \qquad \psi=\psi(r)\,,
\label{gauge3}
\eea
while a convenient Ansatz for the metric takes the form
\beq
ds^2=-{r^2 f(r) \over L^2}dt^2+{L^2 h(r)^2 \over r^2 f(r)} dr^2 -2
C(r) \frac{r^2}{L^2}dt dx+{r^2 \over L^2}B(r) dx^2+{r^2 \over L^2}
dy^2 + {r^2 \over L^2} dz^2\,.
\label{ModTisza4}
\eeq
Since we are interested in finite temperature setups, we will look for solutions
describing a black hole with horizon at $r=r_H$, whose temperature is given
by Eq.~\eqref{eq:tempsf} in the appendix.
The metric contains four independent functions, $f(r),h(r),C(r)$
and $B(r)$. Together with the Ansatz for the gauge field and the
scalar they give rise to a set of seven independent equations for
seven unknowns.
As described in appendix \ref{App:SFF} a two-parameter family of 
superfluid solutions at finite superfluid velocity was found in
\cite{Arean:2010wu}. We parametrize them in terms of their ratio
of temperature over chemical potential $T\over \mu$, and superfluid
velocity along $x$, which 
can be read from the asymptotic values of the gauge field as
(see \cite{Arean:2010wu} and  appendix \ref{App:SFF} for more details)
\be
\zeta=\frac{A_{x,0}}{A_{t,0}}\,.
\ee

As explained in appendix \ref{App:SFF} we find fully back-reacted solutions corresponding to a superfluid 
flow. Since these solutions are obtained in a consistent truncation of string theory \cite{Gubser:2009qm} 
one can, therefore, trust them in a wider regime of validity. In particular, going to very low temperatures,
including the zero temperature limit is completely under control. It is also worth pointing out that the
background we consider is related to a  gravity derivation of the Tisza-Landau model presented 
in~\cite{Sonner:2010yx} which recovers the two-fluid paradigm holographically. Our analysis opens the 
possibility of studying, for example, the behavior of various properties as functions of the normal 
to superfluid fraction for different velocities of the flow.

\subsection{An external particle in holographic superfluid flows}
\label{ssec:SFFlow}

An external probe has already been considered holographically in the context of a superfluid 
by Gubser and Yarom \cite{Gubser:2009qf}. We reproduce the findings presented 
in \cite{Gubser:2009qf} 
and further explore two important generalizations. 
One generalization pertains to considering finite temperatures that are below the critical temperature, 
that is, remaining within the superfluid phase. As the second generalization we consider turning on
a superfluid flow velocity. 

Once we have numerically constructed the superfluid solutions described above,
the drag can be computed by means of Eq.~\eqref{eq:dragsf} with $r_*$ given by the solution
of Eq.~\eqref{u*}.

\begin{figure}[bht]
\centering
	\includegraphics[scale=0.75]{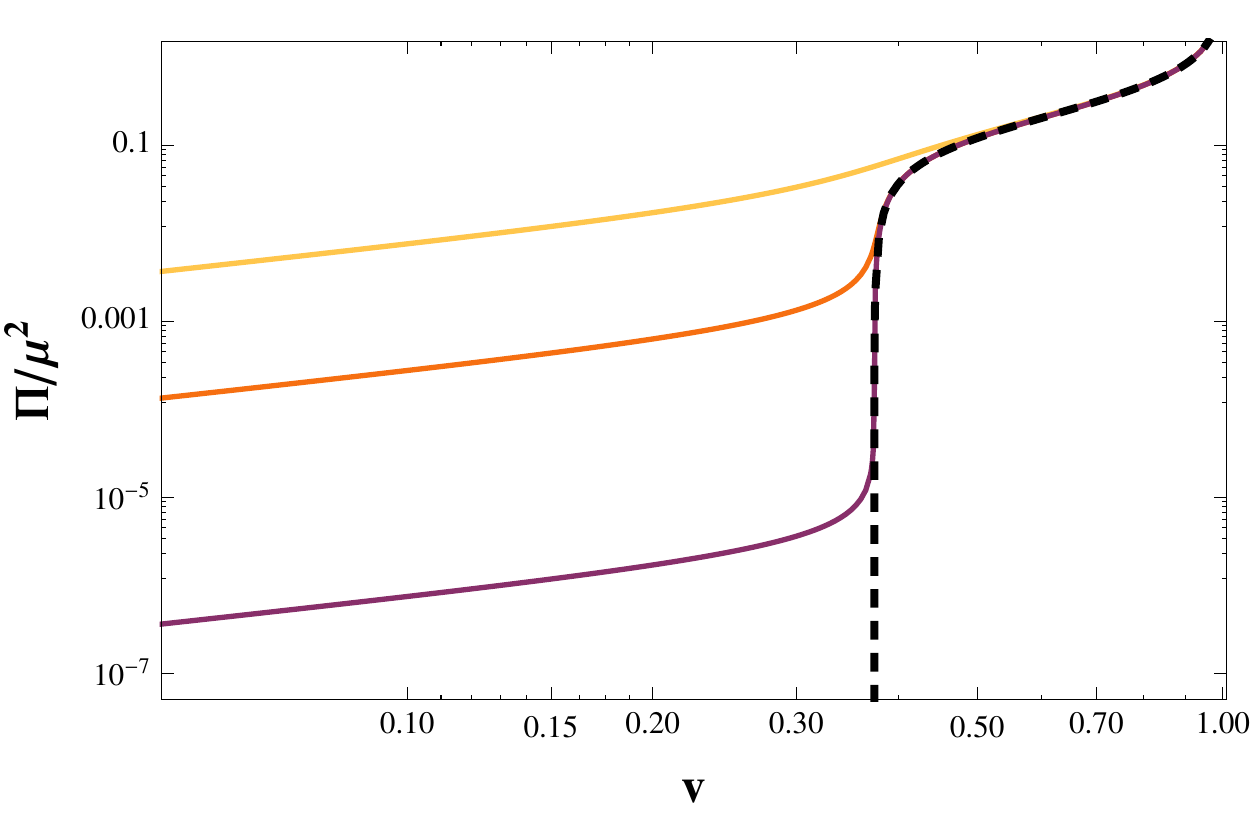}
\captionof{figure}{Drag at zero superfluid velocity: black dashed (T=0),\protect\footnotemark 
purple ($5\times 10^{-3}\,T_c$), orange ($0.11\, T_c$), and  yellow ($0.49\, T_c$).}
	\label{DragSFZero}
\end{figure}

\begin{figure}[bht]
\centering
\includegraphics[width=0.49\textwidth]{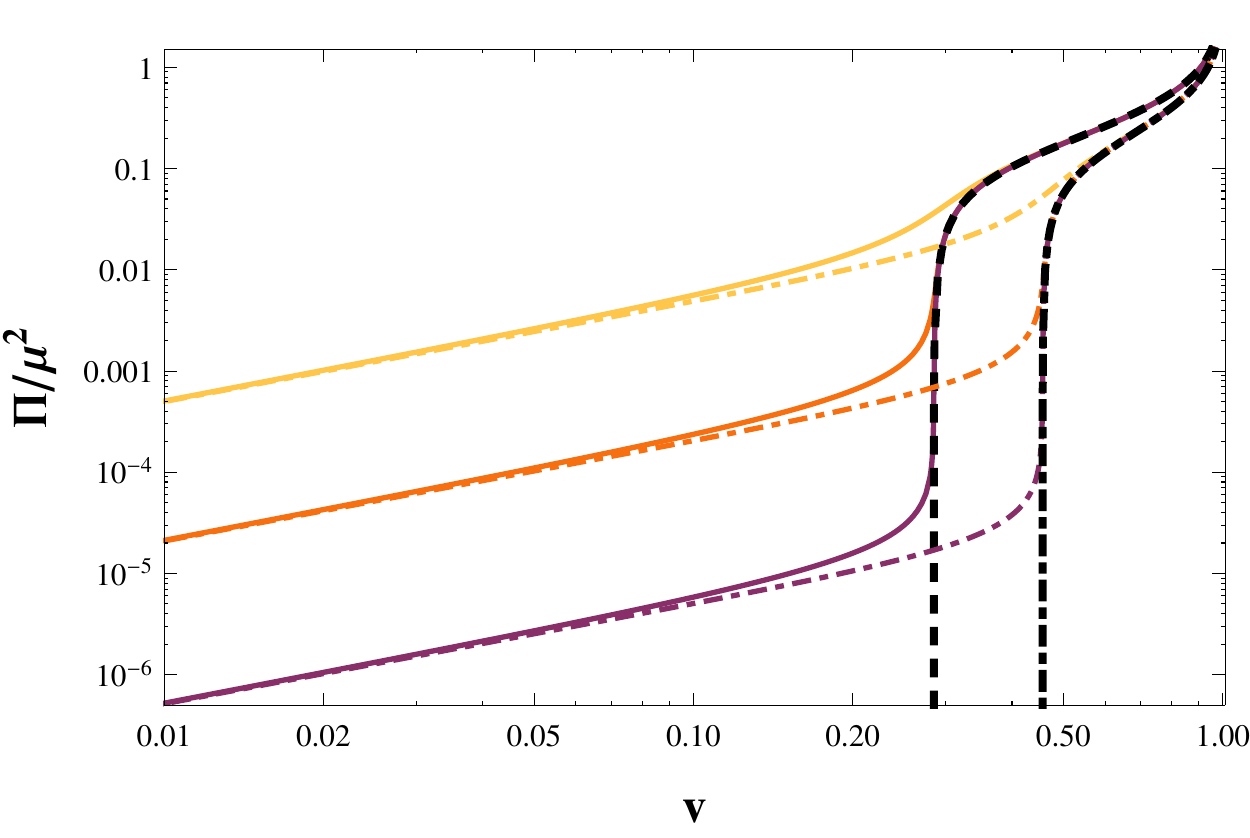}
\includegraphics[width=0.49\textwidth]{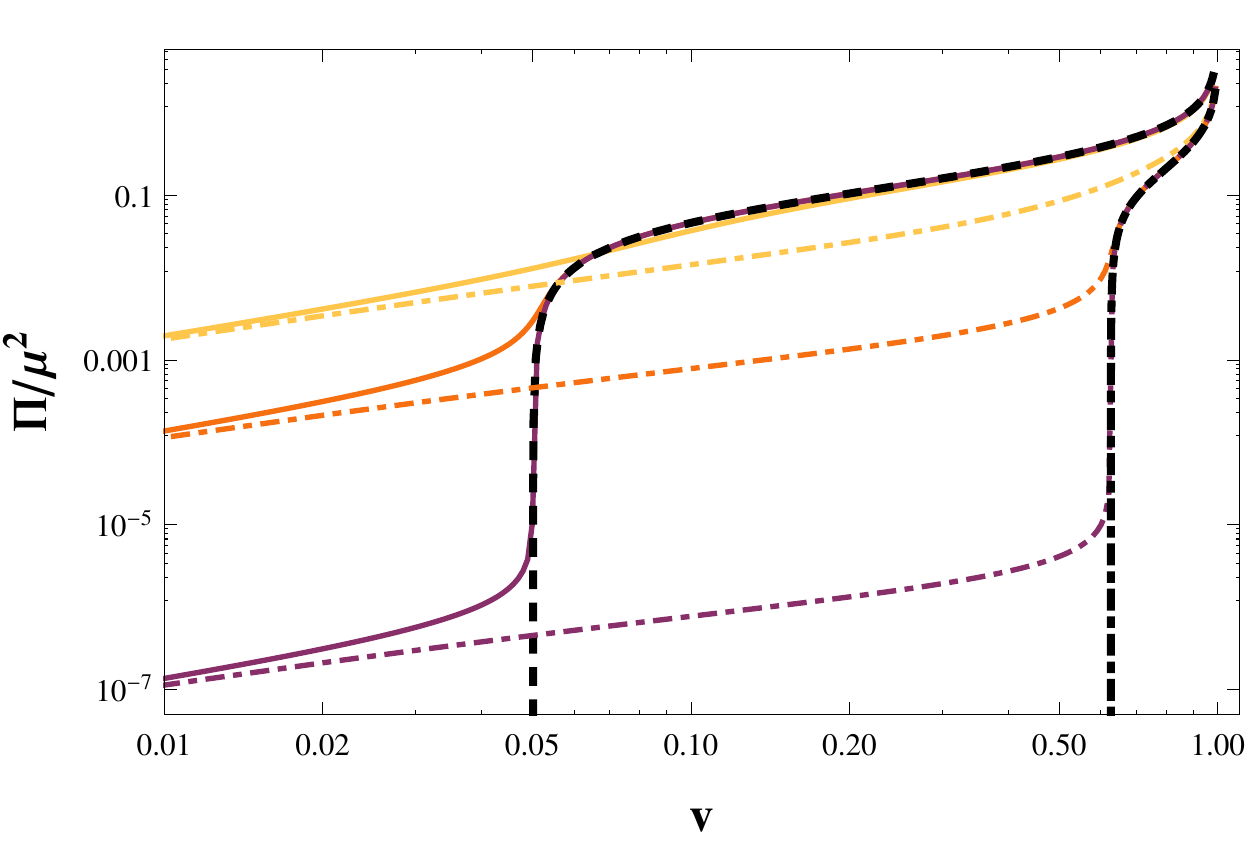}
\captionof{figure}{Left: $\zeta=0.1$. Black dashed  and dot-dashed lines show data at $T=0$,
purple solid and dot-dashed correspond to\protect\footnotemark~ $0.013\, T_c$,
with positive and negative superfluid velocity respectively.
Analogously, orange lines are data at $0.084\, T_c$,
and the yellow ones correspond to  $0.40\, T_c$.
Right: $\zeta=0.33$. Black lines correspond to $T=0$, purple to $T=0.0030\,T_c$, orange to  $T=0.096\,T_c$,
and yellow to $T=0.48\,T_c$. As on the left, dot-dashed lines represent data at $v<0$.}
\label{SmallSFvelocity}
\end{figure}

Figure~\ref{DragSFZero} represents the drag at zero superfluid velocity. The zero temperature limit, 
depicted with a black dashed curve, was discussed in \cite{Gubser:2009qf} and here we essentially 
reproduce those findings.  More important from our point of view is the smooth behavior of the drag 
as we increase the temperature.  For zero temperature a particle experiences no drag below a certain
critical superfluid velocity $v_c\approx0.374$, as established in \cite{Gubser:2009qf}.
Geometrically this is a consequence of the 
zero temperature solution constructed in~\cite{Gubser:2009gp} being a domain wall between two $AdS_5$ 
spaces which is qualitatively different
to solutions with a finite temperature horizon. Figure \ref{DragSFZero} shows how the zero temperature 
behavior is modified as one increases the temperature.
\footnotetext{Here and in Fig.~\ref{SmallSFvelocity} the zero temperature solutions were constructed
as described in~\cite{Arean:2011gz}.}
Notice that as soon as a small temperature is switched on, a black-hole horizon appears in the geometry.
At low enough velocity the endpoint of the string gets close enough to the horizon to recover the analytic  linear
scaling of section~\ref{sec:DragGeneral}. However, if the temperature is low enough, the geometry develops an `intermediate scaling region' around
the horizon where the metric is approximately that of the $AdS_5$ corresponding to the endpoint
of the $T=0$ domain wall. Consequently, for low temperatures, at velocities about $v_c$, the drag
behaves very closely to the $T=0$ case. This is manifest in the behavior of the purple and orange curves
at $v\gtrsim v_c$, and as we will see below in Fig.~\ref{Fig:slopes}, in this region, the drag scales with $v$ as
expected for the $T=0$ case~\cite{Gubser:2009qf}.
As the temperature is raised, the intermediate scaling region disappears from the geometry, and
in Fig.~\ref{DragSFZero}, yellow curve, we arrive at a situation 
which has no memory of the zero temperature limit or the existence of a critical velocity. 
The behavior corresponds strictly 
to a background with finite temperature horizon, as discussed analytically in Section \ref{sec:DragGeneral}.


Figure \ref{SmallSFvelocity} focuses on the effects of switching on a superfluid velocity $\zeta$.
The corresponding finite temperature geometries were constructed in~\cite{Arean:2010wu},
where it was shown that at low temperatures, for velocities below $\zeta_c=v_c$
the metric around the horizon
tends to that of $AdS_5$, exactly as it happened in the $T=0$ case. This picture was confirmed 
in~\cite{Arean:2011gz}, where the zero temperature solutions for $\zeta<\zeta_c$ were constructed, and
shown to be anisotropic domain walls between two $AdS_5$. 
In view of this, for $\zeta<\zeta_c$ one expects a similar behavior of the drag as a function of $v$
to that observed in the $\zeta=0$ case above.
Indeed, the plots in Fig.~\ref{SmallSFvelocity} show the same qualitative
features as Fig.~\ref{DragSFZero},
with the main difference being that the critical velocity $v_c$ is now shifted:
for $\zeta=0.1$ we find it to be $\tilde v_c\approx 0.285$ for positive velocities 
($\hat v_c \approx 0.457 $ for $v<0$),
while for $\zeta=0.33$ it is given by $\tilde v_c\approx 0.050 $ for positive velocities ($\hat v_c =0.63$
for $v<0$).
This behavior is intuitively clear and follows from a simple composition of velocities.\footnote{
Notice that in our units $c=1$, and thus the relativistic composition of velocities must be employed.}
In particular this makes clear that with our conventions $v>0$ corresponds to the particle moving in the
opposite direction to the superfluid flow.
\footnotetext{$T_c$ is the critical temperature at the 
corresponding superfluid velocity $\zeta$, and, as observed in~\cite{Arean:2010wu},
it decreases with increasing $\zeta$.}

\begin{figure}[thb]
\centering
\includegraphics[scale=0.75]{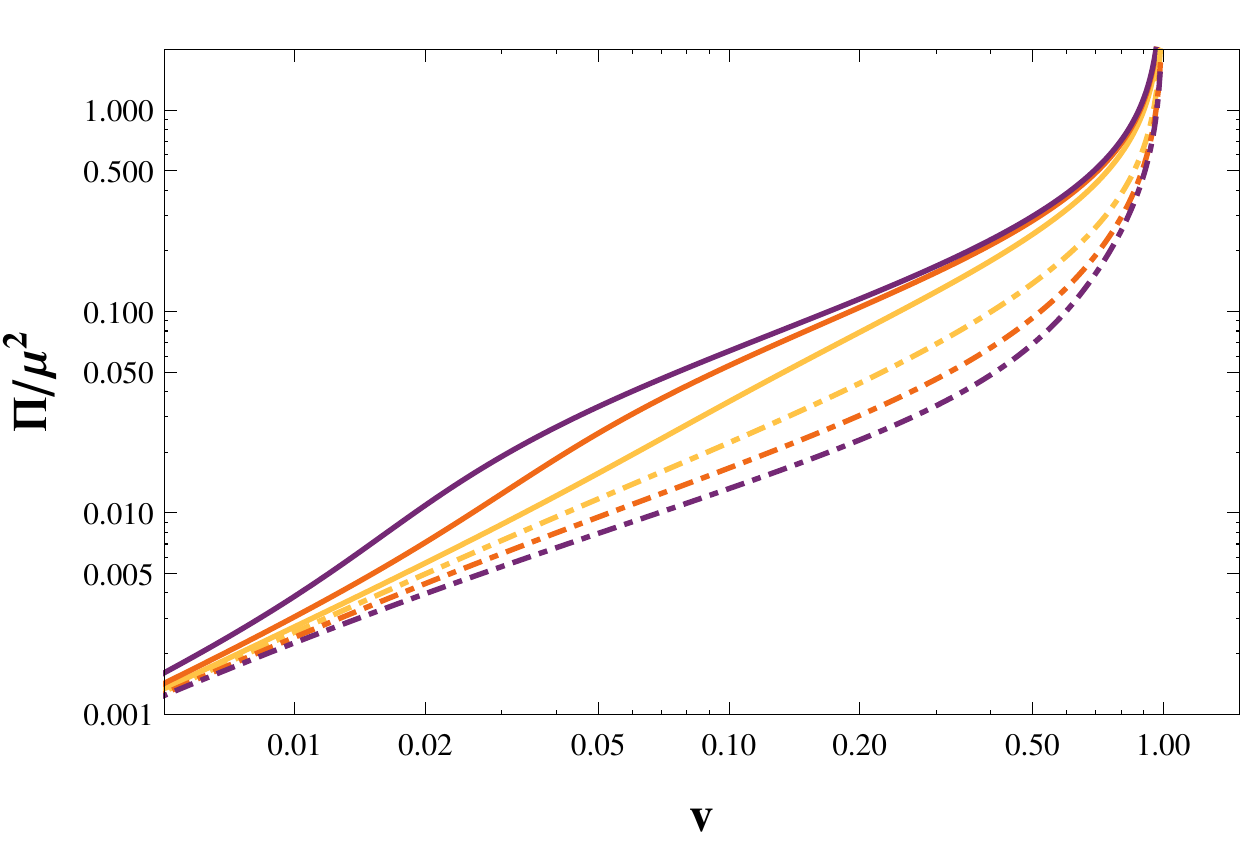}
\captionof{figure}{Large superfluid velocity $\zeta=0.375$. 
For positive velocities: purple solid line ($T=0.16\,T_c$), 
orange solid line ($T=0.30\, T_c$), and yellow solid line ($T=0.62 T_c$).
The dot-dashed lines correspond to negative velocities at the same temperature as their
solid counterparts.}
\label{LargeSFvelocity}
\end{figure}
In Figure \ref{LargeSFvelocity} we explore the regime of large (above critical) superfluid velocity.
As it was observed in~\cite{Arean:2010wu}, at low temperatures and for values of $\zeta>\zeta_c$ 
the superfluid behaves qualitatively differently from the case $\zeta<\zeta_c$. In particular, 
in the dual geometry no $AdS_5$--like region appears around the horizon, and quantities like the value
of the order parameter, or the superfluid fraction, do not converge to the universal
zero temperature value found for $\zeta<\zeta_c$. It was subsequently shown in~\cite{Arean:2011gz}
that if the zero temperature limit of these solutions exists, it cannot be an $AdS_5$ to $AdS_5$
domain wall continuously connected with those found for $\zeta<\zeta_c$. Indeed,  the nature of the zero temperature solution $\zeta>\zeta_c$ remains an open question. In the plot in Fig.~\ref{LargeSFvelocity} we consider the scenario where the superfluid velocity is
slightly above $\zeta_c$, namely $\zeta=0.375$,
and plot the drag for three values of the temperature: $T=0.16\,T_c$ 
(purple lines), $T=0.30\,T_c$ (orange lines), and $T=0.62\,T_c$ (yellow lines).
At first sight, the curves are very different from those at $\zeta<\zeta_c$ of
Fig.~\ref{SmallSFvelocity}, and as expected, there is no sign of the existence of a critical 
velocity $\tilde v_c$.
However, a closer inspection reveals that the curves 
in Fig.~\ref{LargeSFvelocity} are similar to those in Fig.~\ref{SmallSFvelocity} for
$v>\tilde v_c$ (or $v>\hat v_c$ for negative velocities). 
Namely, the external particle in a superfluid at $\zeta>\zeta_c$ experiences
a medium similar to what it encounters in a superfluid 
at $\zeta<\zeta_c$ when moving at $v>\tilde v_c$.
Finally, an interesting feature of Fig.~\ref{LargeSFvelocity} is that
although the magnitude of the drag does not differ notably 
from one curve to another, it is clear that for intermediate and positive values of the velocity, 
the drag is larger for lower temperature (the purple solid curve is above the other two solid curves);
while the opposite occurs for negative velocities (the dot-dashed purple curve falls below the
other two).\footnote{This phenomenon can be observed for the data at $\zeta<\zeta_c$ if one zooms in
on the $v>\tilde v_c$ region (or $v>\hat v_c$ for negative velocities)
and considers temperatures in the range $0.1\, T_c$ -- $0.5\,T_c$.}

Finally, another consequence of Fig.~\ref{DragSFZero}, and ultimately of the presence in the
geometry of first a black hole horizon, and then an approximate $AdS_5$ region, is the 
existence of two different regimes for the drag that could lead to
two different types of velocity statistics.
As said above, for velocities below the critical $\tilde v_c$ (or $\hat v_c$ for $v<0$),
one expects to encounter the linear 
behavior of Section \ref{sec:DragGeneral}. Then, above but close to $\tilde v_c$ (or $\hat v_c$ for $v<0$),
the behavior described in \cite{Gubser:2009qf} is expected.
{\it A priori} these two behaviors (one dictated by the presence of a black hole horizon,
and the other due to an $AdS_5$--like region) are independent, and one could expect to
find two different slopes (for $\log(\Pi)$ versus $v$), and consequently two different
velocity statistics. However, as already noticed in \cite{Gubser:2009qf} for the
zero temperature and zero superfluid velocity $AdS_5$ IR geometry, the IR dimensions
of the fields determining the behavior of the drag are such that again a linear trend
is obtained. 
In Fig. \ref{Fig:slopes} we show how at low temperature our numerics verify 
this prediction for a nonzero $\zeta<\zeta_c$, finding two regions of linear scaling for $\log(\Pi)$.
The black solid line corresponds to the numerical data, and the red dashed lines
to the corresponding logarithmic fits in the regions $v\approx0$ (left panel)
and $ v\gtrsim\tilde v_c$ (right panel). 
As reported in the caption, these fits confirm the expected linear behavior.
\begin{figure}
    \begin{subfigure}[b]{0.49\textwidth}
        \includegraphics[width=\textwidth]{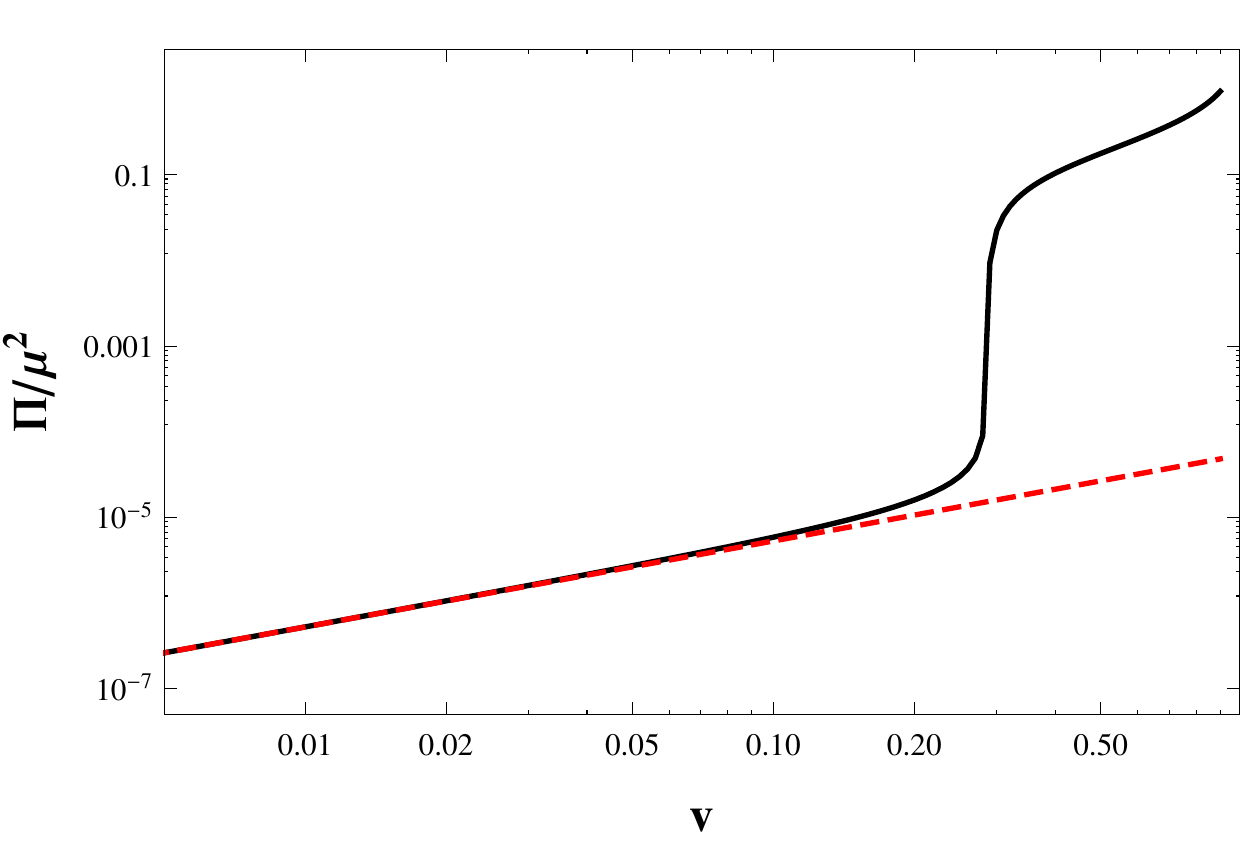}
        \caption{Linear fit (red dashed line) for low velocities at $\zeta=0.1$, $T=0.013 T_c$.
        $\log(\Pi/\mu^2)=1.00 \log(v)-9.84$. We have fitted the data (black solid line) in the 
        region $0.003\leq v<0.0098$.}
        \label{fig:gull}
    \end{subfigure}
    \quad
    \begin{subfigure}[b]{0.49\textwidth}
        \includegraphics[width=\textwidth]{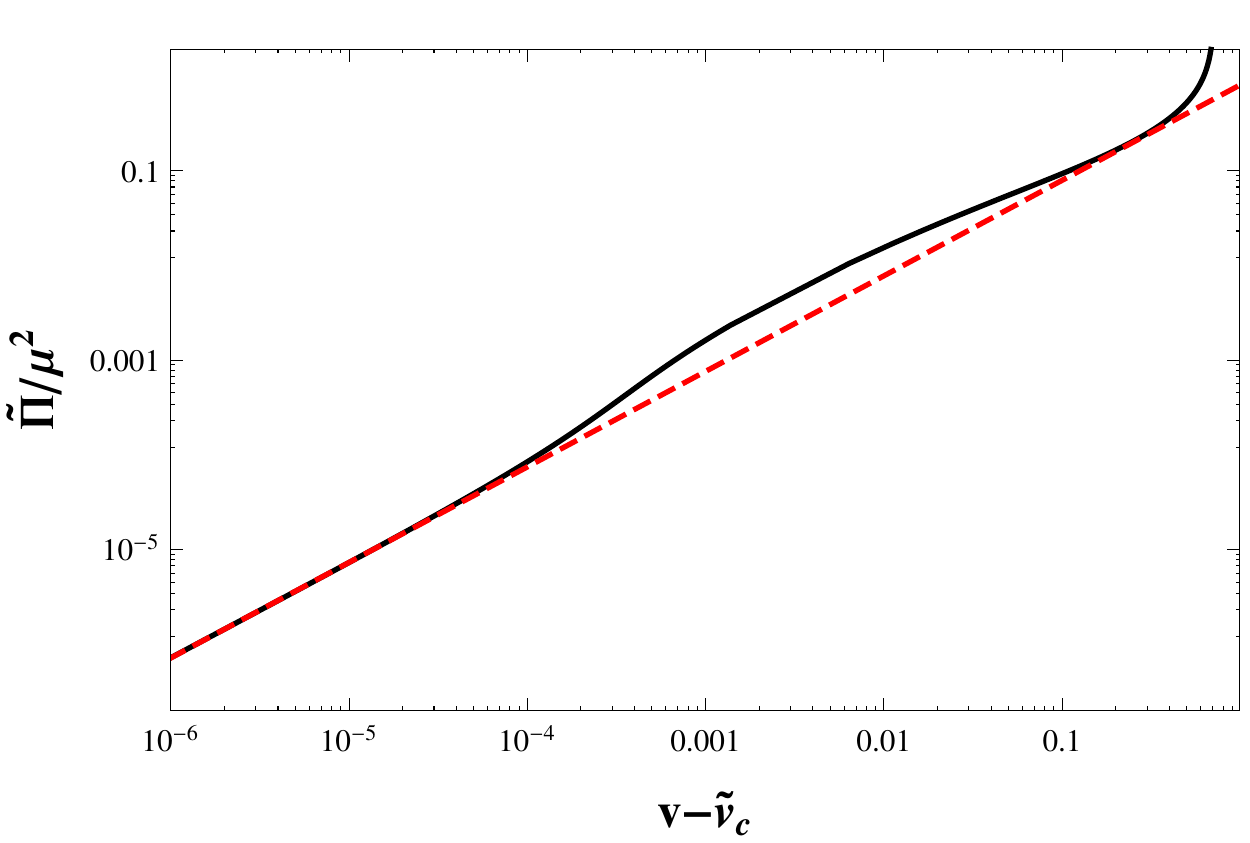}
        \caption{Linear fit (red dashed line) for $v\gtrsim \tilde v_c$ at $\zeta=0.1$, $T=0.013 T_c$.
        $\log(\tilde \Pi/\mu^2)=1.01 \log(v-\tilde v_c)-0.22$. We have fitted the data 
        (black solid line) in the region  $10^{-6}\leq v-\tilde v_c <1.6\times10^{-5}$, and
        defined $\tilde \Pi/\mu^2=\Pi/\mu^2$ at $v=\tilde v_c)$.}
        \label{fig:tiger}
    \end{subfigure}
    ~ 
    \caption{The two slopes for velocities below and above the critical velocity.}\label{Fig:slopes}
\end{figure}

\section{Conclusions}\label{Sec:Conclusions}

 In this manuscript we started by discussing the drag force, and consequently the velocity statistics, 
 of an external heavy particle immersed in a generic holographic fluid. We begun with an analytic 
 discussion of various cases and ultimately found that for dual geometries containing a  finite 
 temperature horizon the velocity statistics goes as $1/v$. Thus, this result is quite universal 
 for non-relativistic motion in fluids that admit a holographic description. We then moved to a 
 more phenomenological approach and by considering Lifshitz geometries determined that a $1/v^3$
 velocity statistics is possible for $z=2$. Finally, in a completely engineering mode, we answered
 the question of what behavior of the metric near the part of the geometry corresponding to the IR,
 leads to $1/v^3$ velocity statistics  which is experimentally observed  for  
hydrogen particles moving in superfluid Helium.

We complemented our general description by numerically discussing two holographic fluids with anisotropies. 
In particular, we have considered the holographic dual of a magnetized plasma where the anisotropy has been 
introduced through the presence of a magnetic field and a holographic superfluid flow where the anisotropy
has been introduced via a velocity in the superfluid flow. Our motivation was to study the effect 
of the anisotropies in the worldsheet horizon, in particular, to explore the extent to which the 
approximations made in the analytic exploration hold more generally. 

In the case of a magnetized quark gluon plasma we studied the velocity statistics for a particle moving parallel and perpendicular to the magnetic field. We found that the magnetic field has a very small effect on the velocity statistics. In the non-relativistic regime we found $1/v$ behavior but we detected small deviations as a function of the magnetic field for intermediate velocities. The case of a very strong magnetic field deserves some comments, as we observed that the drag force for movement in the direction of the field grows only up to a certain limit for each velocity as the magnetic field goes to very large intensities. This is in contrast with the behavior for perpendicular movement, case in which for any given velocity the drag force grows without a bound as the intensity of the field gets larger. This is of particular relevance as it confirms the expectation that a field theory in the presence of a strong magnetic field should behave like a lower dimensional theory eliminating the directions perpendicular to the magnetic field. This phenomenon was also reported in \cite{Li:2016bbh, Mamo:2016xco}, where the geometry $AdS_3\times T^2$ that we use as the near horizon limit, see appendix \ref{App:MagBack} for details, was extended all the way to the boundary, hence not approaching $AdS_5$, but capturing the physics of large $B$.

In the case of  a superfluid with vanishing flow velocity, we reproduced the intuitive results of  \cite{Gubser:2009qf}. We establish, as in \cite{Gubser:2009qf}, that there is a critical velocity below which there is no drag. For finite temperature solutions we found that there is a drag that for small velocities follows the universal behavior discussed analytically. We studied the effect of a small and a large, compared to the critical velocity, superfluid velocities. Although, generically the velocity statistics for velocities above the critical velocity can be different from $1/v$ we found that in the particular top-down model we considered the statistics is compatible with $1/v$ with minor deviations.

Let us conclude by clarifying an important point of our survey of holographic fluids. 
The standard notion of turbulence has as an important component the presence of vortices. Holography has already produced some interesting setups for turbulence \cite{Adams:2013vsa,Green:2013zba}. In particular, a model for turbulence in a superfluid, that is,  quantum turbulence was presented \cite{Adams:2012pj}. It would be natural to consider  an external heavy probe in those setups. However, our motivation in this manuscript  is slightly different. Namely, normally superfluid turbulence is a chaotic motion of well-determined and well-separated vortex filaments; we are, however, interested in a different kind of potential superfluid turbulence having to do more with the strongly coupled nature of the interactions in the medium.  Indeed, wave turbulence is less familiar than vortex turbulence, it occurs in systems of strongly interacting nonlinear waves; it can lead to flows across length and frequency scales just as those of vortex turbulence. It has been established in systems described by the Gross-Pitaevskii  
(nonlinear Schr\"odinger) equation;  see the review \cite{Kolmakov25032014}. 
In an AdS/CFT motivated context, there have appeared  interesting new results in wave turbulence
in spherically symmetric situations in AdS spaces. For example, weak turbulence during gravitational 
collapse in  AdS discovered in \cite{Bizon:2011gg} was later shown to be quantitatively described by 
a Kolmogorov-Zakharov spectrum \cite{deOliveira:2012dt}.  These types of results justified our exploring
holographic  fluids with no vortices at all using external heavy particles.

There are a number of questions that our investigation opens. For example, one of the interesting questions that we hope to address in the future pertains the crossover between classical and quantum superfluid turbulence. This is a question that has been tackled using wave turbulence methods in \cite{PhysRevB.76.024520}. In the model presented in section \ref{Sec:SFFlow},  the temperature controls the superfluid fraction and one could study how the velocity of the flow affects its properties in a concrete holographic setup which is relevant for strongly coupled superfluids.

\section*{Acknowledgments}
We would like to thank C\'esar Terrero-Escalante for discussions and participation in 
the gestation stages of this project. We also thank Dori Reichmann for extensive discussions 
a few years ago.  We are thankful to Amos Yarom for various comments and clarifications. 
D. A. thanks the University of Michigan, Ann Arbor  and the ICTP, Trieste for hospitality during the 
course of this work; and the FRont Of pro-Galician Scientists for unconditional support.
L. P-Z. thanks L. P. for warm hospitality while visiting Cinvestav. 
The work of D.A. is supported by the German-Israeli Foundation (GIF), grant 1156.
L. P. and M. V. acknowledge
partial support from DGAPA-UNAM Grant No. IN113115.

\appendix

\section{Gravity dual of a magnetized quark-gluon plasma}\label{App:MagBack}
As  already mentioned in the main text, the gravitational background that we used as dual to a magnetized quark-gluon plasma is described by a metric of the form
\begin{equation}
ds^2=-U(r)dt^2+V(r)(dx^2+dy^2)+W(r)dz^2+\frac{dr^2}{U(r)},\label{metansA}
\end{equation}
and a field strength given by
\begin{equation}
F=B dx \wedge dy.\label{FansA}
\end{equation}

Provided the form of (\ref{metansA}) and the exclusive dependence on $r$ of the metric potentials, $F$ 
verifies Maxwell equations regardless of the specific shape of $U, V$ and $W$, which in 
turn are determined by Einstein equations
\begin{align}\label{eqfond}\nonumber
 2 W(r)^2 \,\bigg[4 B^2  +V(r)\, \Big(U'(r)\, V'(r)+U(r) \, V''(r)\Big)\bigg]\, -\, V(r)\, W(r)\,\, \bigg[2 V(r) \\ \nonumber
 \times\Big(U'(r) W'(r)+U(r) W''(r)\Big)+U(r) V'(r) W'(r)\bigg]+U(r) V(r)^2 W'(r)^2 &= 0,\\
  4 V(r) W(r)^2 V''(r)-2 W(r)^2 V'(r)^2-V(r)^2\Big(W'(r)^2-2 W(r) W''(r)\Big) &= 0,  \\ \nonumber
  W(r) \bigg[-8 B^2+6 V(r)^2 \Big(U''(r)-8\Big)+6 V(r) U'(r) V'(r)\bigg]+3 V(r)^2 U'(r) W'(r) &= 0,\\  \nonumber
 W(r) \Big(4 B^2+2 V(r) U'(r) V'(r)+U(r) V'(r)^2 -24 V(r)^2\Big)+V(r) W'(r)\\ \nonumber
\times  \Big(V(r) U'(r)+2 U(r) V'(r)\Big) &= 0.
\end{align}

Out of the four equations (\ref{eqfond}) only the first three are dynamical, and the fourth is a 
constraint that the dynamical equations will keep true for all times as long as it is set to be 
satisfied at some initial time. The main effect that this constrain has is that it reduces the 
space of solutions of (\ref{eqfond}) to be four dimensional, whereas usually a set of three second 
order differential equations for three functions would lead to a six dimensional space of solutions.

There are two analytic solutions to equations (\ref{eqfond}) which are of particular interest to us.
One of them is the direct product of a BTZ black hole and a two dimensional space that 
in \cite{D'Hoker:2009mm} was taken to be $T^2$. This solution is described by inserting
\begin{equation} \label{btz}
U_{BTZ}(r)=3(r^2-r_h^2),\;\;\; V_{BTZ}(r)=\frac{B}{\sqrt{3}}\;\;\; \mathrm{and}\;\;\; W_{BTZ}(r)=3r^2,
\end{equation}
in (\ref{metansA}), and it becomes apparent that the solution has an horizon at $r_h$. One important result of  \cite{D'Hoker:2009mm} is the  successful construction of  a numerical solution that interpolates between (\ref{btz}) for $r\to r_h$ and  $AdS_5$ for $r\to \infty$. It will be useful below to notice now that this solution can be written exactly as the series expansion
\bea
U_{BTZ}(r)&=&6 r_h (r-r_h)+3(r-r_h)^2, \nonumber \\
V_{BTZ}(r)&=&\frac{ B }{\sqrt{3}}, \label{BTZhor}\\
W_{BTZ}(r)&=&3 {r_h}^2+6 r_h (r-r_h)+3(r-r_h)^2. \nonumber
\eea

The other analytic solution that is relevant to us is given by plugging
\begin{align}
\label{bb}
U_{BB}(r) &=(r+\frac{r_h}{2})^2(1-\frac{(\frac{3}{2}r_h)^4}{(r+\frac{r_h}{2})^4}),\nonumber\\
V_{BB}(r) &=\frac{4V_0}{9r_h^2}(r+\frac{r_h}{2})^2,\\
W_{BB}(r) &=\frac{4}{3}(r+\frac{r_h}{2})^2, \nonumber
\end{align}
in (\ref{metansA}) and setting $B=0$. This is nothing else than a scaled version of the black
D3-brane solution, as it becomes apparent by shifting the radial coordinate as $\tilde{r}=r+\frac{r_h}{2}$,
leaving the horizon at $\tilde{r}=\frac{3}{2}r_h$. The reason to shift the black brane solution in this
way is so that its series expansion around $r_h$ can be written as
\bea
U_{BB}(r)&=&6 r_h (r-r_h)-2 (r-r_h)^2+\frac{8 }{3 r_h}(r-r_h)^3+{\cal O}(4) \nonumber \\
V_{BB}(r)&=&V_0+\frac{4 V_0 }{3 r_h}(r-r_h)+\frac{4 V_0 }{9 r_h^2}(r-r_h)^2, \label{BBP}\\
W_{BB}(r)&=&3 r_h^2+4 r_h (r-r_h)+\frac{4}{3} (r-r_h)^2, \nonumber
\eea
where the only series that does not terminate where indicated is that of $U$, which contains terms of 
fourth order and higher, that are of course determined, but their details are not relevant for the 
present discussion. Writing the expansion of the black brane as (\ref{BBP}) explicitly shows that it 
coincides with (\ref{BTZhor}) up to first order in the series for $U$ and up to zeroth order in the
series for $W$.

We have already mentioned that the space of solutions to equations (\ref{eqfond}) is four dimensional, 
which in the context of their construction as power series, means that only four of the coefficients are 
independent while the rest of them are determined through the field equations once the relevant four are 
fixed. In this light we see that the series expansion
\bea
U_P(r)&=&0\,\, (r-r_h)^0 + 6 r_h (r-r_h)+\sum _{\text{i}=2}^\infty U_i(r-r_h)^i, \nonumber \\
V_P(r)&=&\sum _{\text{i}=0}^\infty V_i(r-r_h)^i, \label{BTZP}\\
W_P(r)&=&3 r_h^2\,\, (r-r_h)^0 +\sum _{\text{i}=1}^\infty W_i(r-r_h)^i, \nonumber
\eea
has already three of the free coefficients fixed while still accommodates both, (\ref{BTZhor}) and 
(\ref{BBP}), enabling us to use $V_0$ as the fourth parameter whose value will determine the particular
solution.

To recover (\ref{BTZhor}) from (\ref{BTZP}), $V_0$ has to be set equal to $\frac{ B }{\sqrt{3}}$, since
only for this value of $V_0$ the field equations force the series to remain finite and terminate exactly 
as in (\ref{BTZhor}). Notice that in this case, for the metric not to be degenerate $B$ cannot vanish.

To recover (\ref{BBP}) from (\ref{BTZP}), $V_0$ is free to take any positive value, and just as long as
$B=0$, the field equations will determine the series to be exactly as (\ref{BBP}). The freedom in $V_0$ 
is just the freedom to rescale the $x$ and $y$ directions by a constant factor.

At first sight, it would seem as if we have three independent parameters, $r_h, B$ and $V_0$ to specify
the gravitational background, but a more careful examination shows first that as far as $B$ and $V_0$ 
are concerned, the only relevant quantity is $B/V_0$, and any independent modification of these two 
parameters that does not change their ratio will correspond to the same field intensity in the field 
theory side. The previous claim is true by construction, since the magnetic field is a 2-form, whose
magnitude square in the field theory side is $\langle F,F \rangle=B^2/{V_\infty}^2$,
with $V_\infty\equiv V(r\rightarrow\infty)$. Since $V_0$ dictates the amplitude of the gravitational
potential $V$ obtained numerically, only the value of the ratio $B/V_0$ modifies the intensity of the
physical field $b\equiv B/V_\infty$ in the field theory side. The last detail we need to indicate is 
that $b$ should be measured in units of $T^2$, being this the one scale we have at hand in the field 
theory, and so our results are reported in terms of the dimensionless quantity $b/T^2$, a magnitude 
that can be established when discussing real world experiment. The temperature is written in terms of
$r_h$ as $T=\frac{3r_h}{2\pi}$, so from the three apparent parameters, all the results should be 
reported in terms of only one quantity, which is $b/T^2$.

We have not been able to find analytic solutions to (\ref{eqfond}) for values of $B/V_0$ other than 0
and $\sqrt{3}$, but we can numerically generate a whole one parameter family of solutions whose members 
are labeled by the value of $B/V_0$. To do this we have to integrate (\ref{eqfond}) numerically starting
at a very small distance $\epsilon$ from $r_h$, where we set the conditions for the metric potentials
$U,V$ and $W$ close to the horizon by equating them and their derivatives to the result of evaluating
(\ref{BTZP}) at $r=r_h+\epsilon$. In this way each value of $B/V_0$ that we feed in the numerical 
construction through (\ref{BTZP}) and (\ref{eqfond}) will generate one member of the one parameter 
family of solutions mentioned before. The way in which we wrote the expansion close to the horizon
was also thought so that, all members of the family of solutions generated by using this expansion
share the same $T=\frac{3r_h}{2\pi}$, associated by the regularity of the Euclidean continuation of
any member of the family.

By seeing through the procedure described in the previous paragraph, we can appreciate that for
$B/V_0\ll 1$ the solutions will look very much like that of a black brane without a magnetic
field, and we will confirm that any results that we find through the gauge/gravity correspondence 
will not depart drastically from those found for the quark gluon plasma without magnetic field. We 
also see that for $B/V_0$ close to $\sqrt{3}$, regardless of how close, just as long as it is not 
equal, the solution will approach (\ref{btz}) near to the horizon, and will asymptote to $AdS_5$ 
for $r\to \infty$.

We close this appendix by noticing that, given the nature of equations (\ref{eqfond}), the 
integration procedure described above will in general result in 
$U(r)\rightarrow r^2$, $V(r)\rightarrow C_1 r^2$ and $W(r)\rightarrow C_2 r^2$ for some constants 
different from the unit as $r$ approaches the boundary. To have a background that is truly asymptotically
AdS, we notice that equations (\ref{eqfond}) are invariant under scalings of $W$. Also, (\ref{eqfond})
are homogeneous in $V$ and $B$ combined, and therefore invariant under simultaneous scalings of $V$ and 
$B$. In practice we use this to first compute a solution using the integration method described earlier 
in this appendix, and once we have obtained a numerical solution, we divide $V$ and $B$ simultaneously
by $C_1$ and $W$ by $C_2$. The metric thus obtained is asymptotically AdS and has $b=B$.

\subsection{More details of the drag in a magnetized quark-gluon plasma}
In this appendix we present more details on the drag  on a external particle moving in a magnetized 
quark-gluon plasma.
\begin{figure}[htb!]
  \centering
  \includegraphics[width=.7\linewidth]{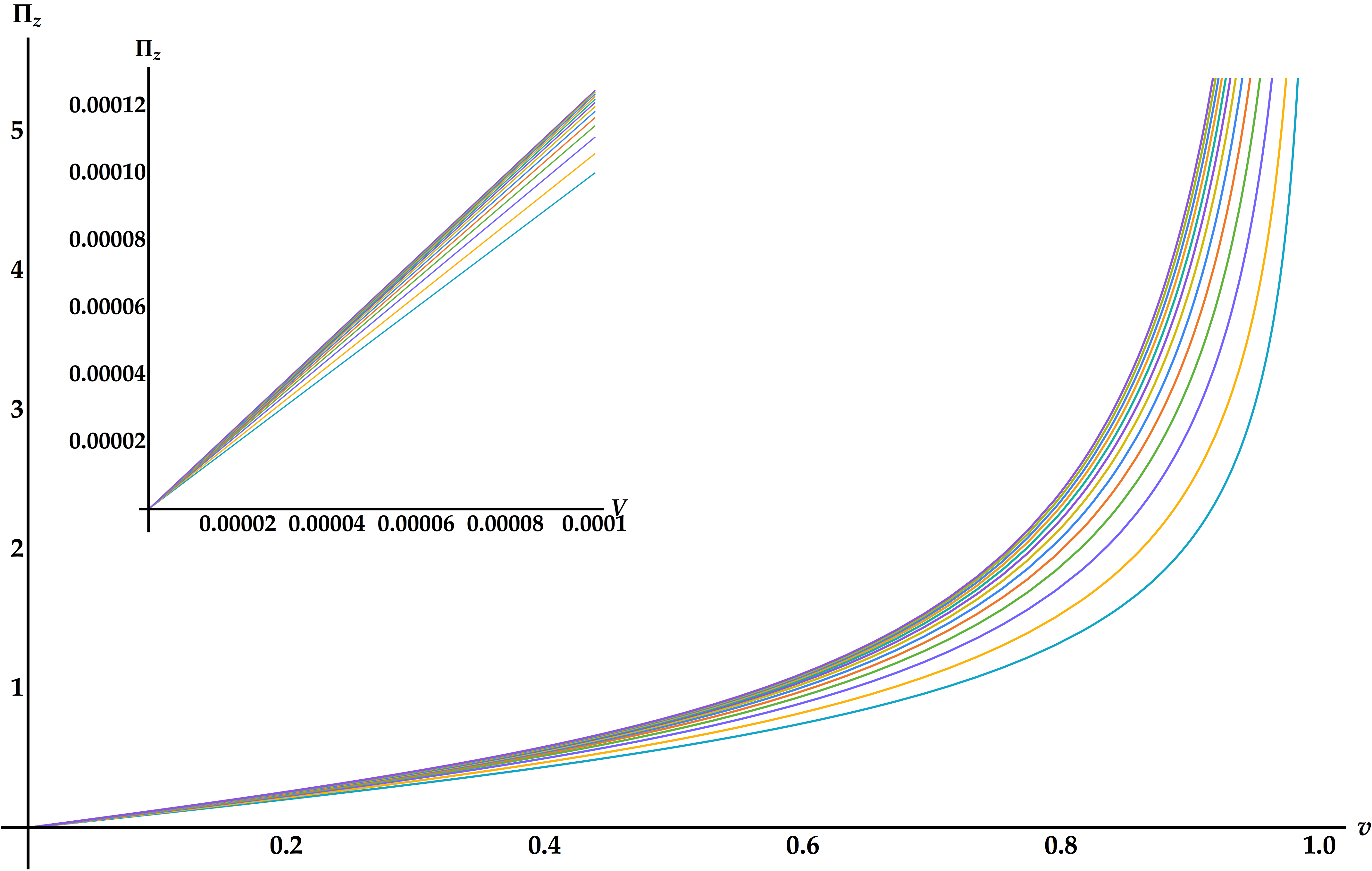} 
  \captionof{figure}{$\Pi_z$ as function of the velocity for $b/T^2=0, 14.74, 30.31, 47.88, 67.14, 87.87, 109.89, 133.06, 157.27, 182.44, 208.50, 235.39, 263.06$ with larger values of $b/T^2$ corresponding with higher lines. Notice that the lowest line demonstrates the consistency of our numerical results with previous analytic findings. The inset shows the behavior of the same plots for velocities very small compared to 1.}
  \label{PzV}
\end{figure}
\begin{figure}[htb!]
  \centering
  \includegraphics[width=.7\linewidth]{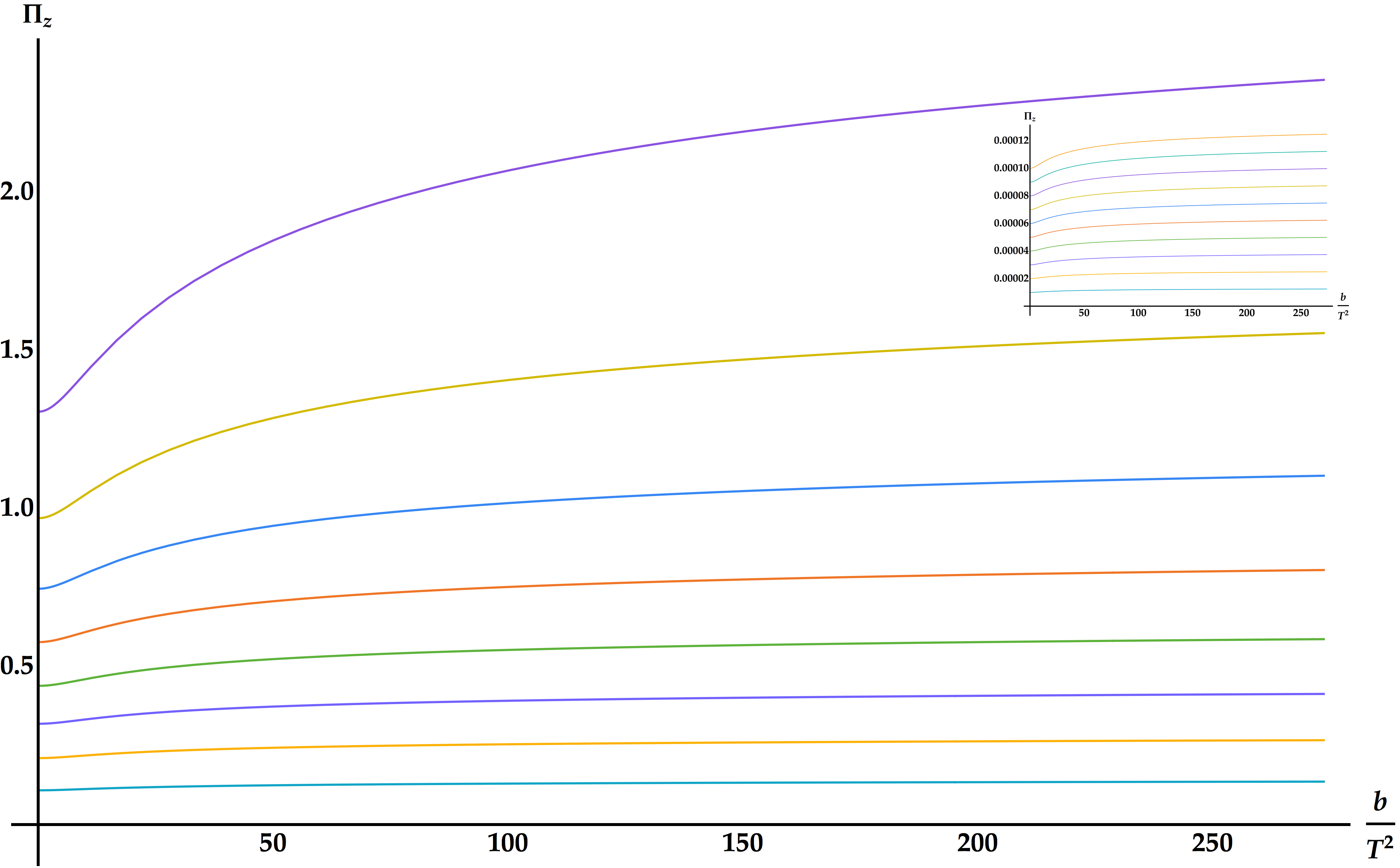} 
\captionof{figure}{$\Pi_z$ as function of $b/T^2$ for values of the velocity 27/250, 103/500, 38/125, 201/500, 1/2, 299/500, 87/125, 397/500, 223/250, 99/100. The inset shows the plots for velocities 1/100000, 1/50000, 3/100000, 1/25000, 1/20000, 3/50000, 7/100000, 1/12500, 9/100000, 1/10000. Larger values of the velocity correspond with higher lines.}
  \label{PzB}
\end{figure}
In Fig. (\ref{PzV}) we present the value of the drag without normalization to the zero magnetic 
field answer.  
Figure (\ref{PzB}) displays the drag as a function of the magnetic field.

\begin{figure}[htb]
  \centering
  \includegraphics[width=.7\linewidth]{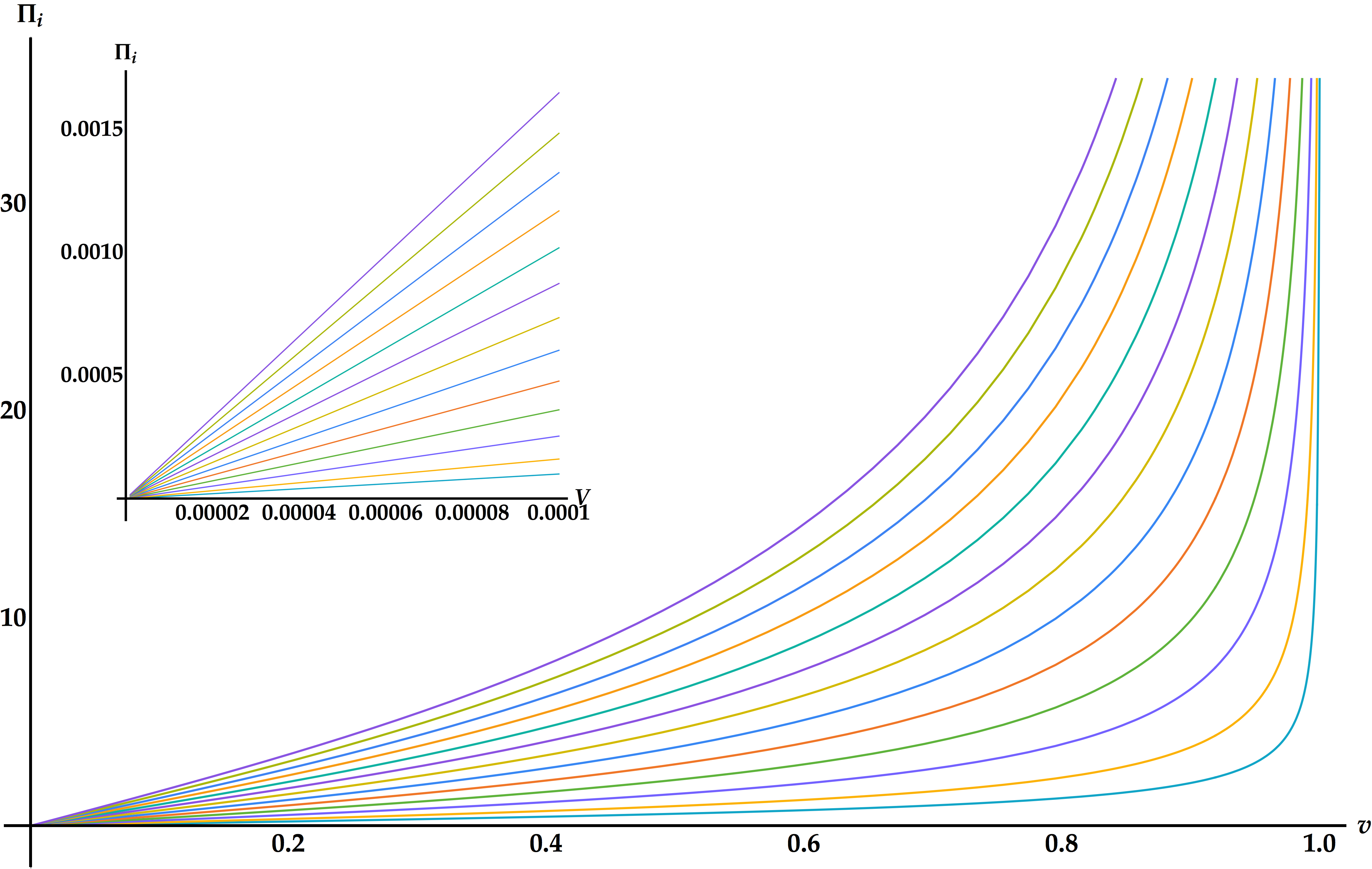} 
  \captionof{figure}{$\Pi_{i=x,y}$ as a function of  velocity, $v$, for $b/T^2=0, 14.74, 30.31, 47.88, 67.14, 87.87, 109.89, 133.06, 157.27, 182.44, 208.50, 235.39, 263.06$ with larger values of $b/T^2$ corresponding with higher lines. Notice that the lowest line demonstrates the consistency of our numerical results with previous analytic findings. The inset shows the behavior of the same plots for velocities very small compared to 1.}
  \label{PxyV}
\end{figure}

\begin{figure}[h!]
\begin{center}
	\includegraphics[width=.7\linewidth]{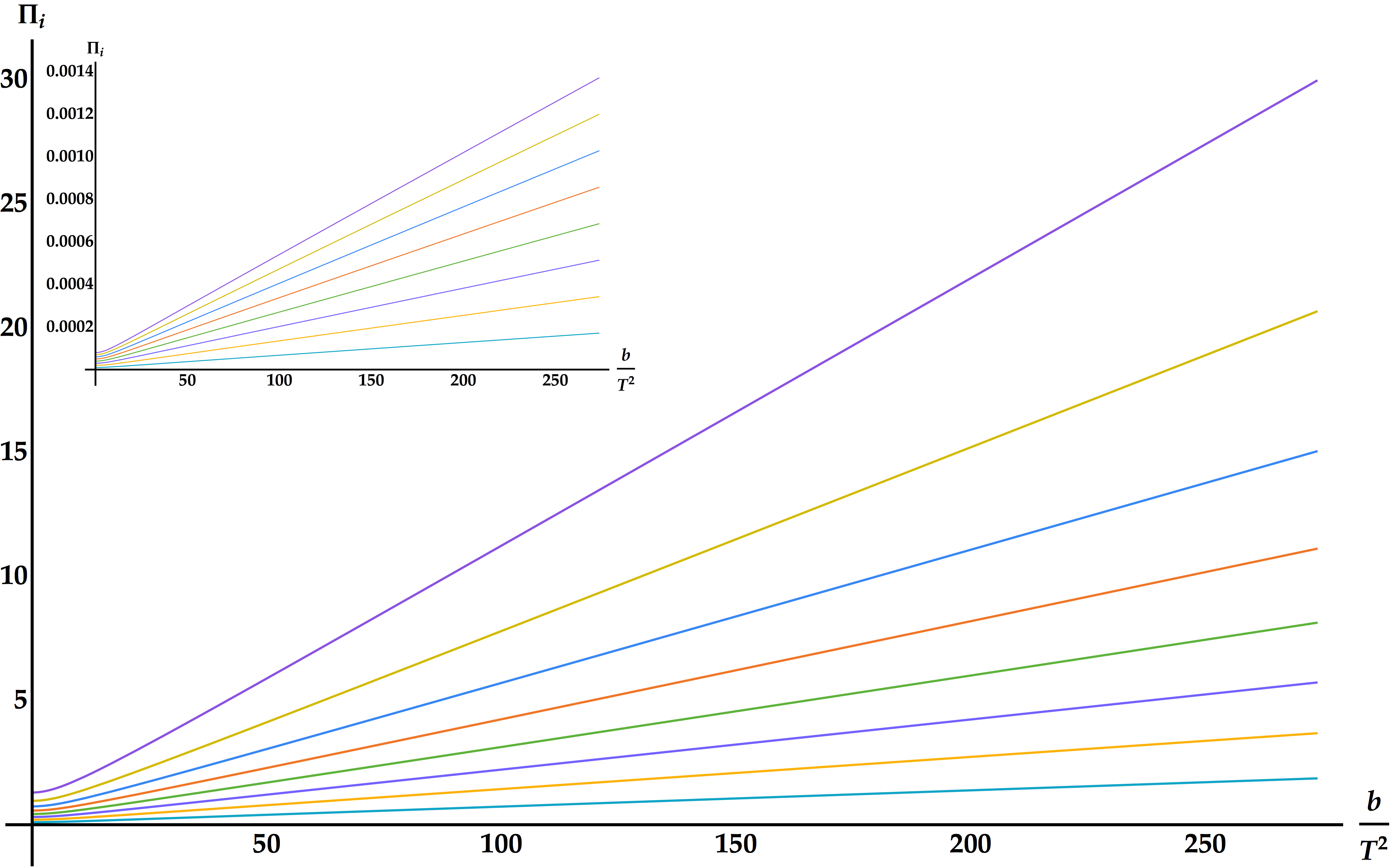} 
	\captionof{figure}{$\Pi_{i=x,y}$ as function of $b/T^2$ for values of the velocity 27/250, 103/500, 38/125, 201/500, 1/2, 299/500, 87/125, 397/500, 223/250, 99/100. The inset shows the plots for velocities 1/100000, 1/50000, 3/100000, 1/25000, 1/20000, 3/50000, 7/100000, 1/12500, 9/100000, 1/10000. Larger values of the velocity correspond with higher lines.}
\label{PxyB}
\end{center}
\end{figure}

In figures (\ref{PxyV}) and (\ref{PxyB}) we present the dependence of the drag  perpendicular to 
the magnetic field as a function of the velocity of the particle and of the magnetic field, respectively. 

\newpage



\section{IIB Holographic superfluid flows}\label{App:SFF}

In this appendix we provide some details about the 
construction of   
the superfluid solutions considered in section~\ref{Sec:SFFlow}. 
More details can be found 
in the original work \cite{Arean:2010wu} 
from which we heavily borrow in an attempt to make our presentation somehow self contained.

An important observation is that our setup possesses several scaling symmetries
which simplify the task of finding a solution.
First, the ambiguity of units for time $t$ and distance $x$ translates into 
two scaling symmetries of the equations of motion
\bea && t \rightarrow
t/\mbox{a}\;, \quad f \rightarrow \mbox{a}^2 f\;, \quad h\rightarrow
\mbox{a} \,h\;, \quad C \rightarrow \mbox{a} \,C\;, \quad A_t
\rightarrow \mbox{a} \,A_t\,,
\label{scale1B} \\
&& x \rightarrow x/\mbox{b}\;, \quad B \rightarrow \mbox{b}^2 B\;,
\quad C \rightarrow \mbox{b}\, C\;, \quad A_x \rightarrow \mbox{b}\,
A_x\,.\label{scale2B}
\eea 
Indeed it
is easy to check that these are symmetries of the action and thus of the equations
of motion. There are two further scaling symmetries of the system 
\bea\label{scale3B} 
&& (r, t, x, y, z, L) \rightarrow \alpha
(r, t, x, y, z, L)\;, \quad (A_t, A_x) \rightarrow (A_t, A_x)/\alpha\,,
\\
\label{scale4B}
&& r \rightarrow \beta r\;, \quad (t, x, y, z) \rightarrow (t, x, y,
z)/\beta\;, \quad (A_t, A_x) \rightarrow \beta (A_t, A_x)\,.  
\eea
The first scaling alters the metric by a factor $\alpha^2$ while leaving the gauge
field unchanged. Its effect amounts to an overall rescaling of the action
\eqref{IIBac} by a constant factor $\alpha^2$, leaving the equations of motion
unaffected. The second scaling corresponds to the usual holographic renormalization
group operation in AdS, and it leaves invariant the metric, gauge field, and equations
of motion.
We will use the symmetries (\ref{scale3B}, \ref{scale4B}) to scale the horizon radius $r_H$ and the AdS
radius $L$ to unity.

We shall start the characterization of the solutions relevant to describe
a superfluid by 
considering the near horizon ($r=r_H$) region and expanding the several fields $\Phi$
in a Taylor series as 
\beq
\label{expH}
\Phi = \Phi_0^H + \Phi_1^H (r - r_H) + \dots\,.
\eeq 
Requiring
regularity of the solution at the horizon amounts to setting some
specific coefficients to zero. To linear order in $(r-r_H)$, the
expansion at the horizon takes the form
\bea
f&=&f_1^H(r-r_H)+... \label{start}\\ h&=&h_0^H+h_1^H(r-r_H)+... \\
B&=&B_0^H+B_1^H(r-r_H)+...\\ C&=&C_1^H (r-r_H)+... \\
A_t&=&A_{t,1}^{H}(r-r_H)+... \\
A_x&=&A_{x,0}^{H}+A_{x,1}^{H}(r-r_H)+...\\
\psi&=&\psi_0^H+\psi_1^H(r-r_H)+...\,.
\label{end}
\eea 
Regularity at the horizon implies that $f_0^H,
C_0^H$ and $\phi_0^H$ should vanish.  Imposing the equations of motion
has the effect of putting further constraints on various coefficients, all of which are 
ultimately determined by a small set of independent horizon data. 
It turns out that the coefficients can all be determined
in terms of six independent data 
\bea 
(h_0^H, B_0^H, C_1^H, A_{t,1}^H, A_{x,0}^H, \psi_0^H)\,.
\label{horizondata5d}
\eea
Therefore, the solutions that we find by integrating
from the horizon form a six-parameter family. All other
coefficients are functions of these ones.  For example, one such relation  is 
\bea
f_1^H=(h_0^H)^2\Big({9 \over 4}+2 \cosh{\psi_0^H}-{\cosh(2\psi_0^H)
\over 4}\Big)-\frac{2 (A_{t,1}^H)^2}{9}\,.  
\eea 
One should next
integrate the solution from the horizon out to the boundary
($r\rightarrow \infty$), starting with the free horizon data
(\ref{horizondata5d}). It is crucial to consider a suitable Ansatz 
for the asymptotics of the fields at the boundary.
In five dimensions that asymptotic expansion is subtle since,
as already noticed, the mass of the
scalar is such that there is a non-normalizable mode. 
To accommodate a
generic solution resulting from integration from the horizon one
then needs consider a boundary expansion where the non-normalizable 
mode of the scalar is turned on.
That non-normalizable mode triggers further
logarithmic terms in the asymptotic expansion, forcing one to keep
track of them too. A combined series expansion
in both $1/r^n$ and $\log r/r^m$ of the form
\bea \Phi = \sum_{n=0}^{\infty}
\Phi_n\,\frac{1}{r^n} + \sum_{m=0}^{\infty} \Phi^l_m \,\frac{\log
r}{r^{m}}
\,,
\eea 
does the job.

Integrating from the horizon and employing a shooting technique 
we select the solutions which match our requirements in the UV.
These are, on one hand, the vanishing of the leading falloff of the scalar field
such that the breaking of the $U(1)$ is spontaneous; and on the other,
the metric becoming asymptotically $AdS_5$ in the UV.

It was found in \cite{Arean:2010wu} that the following asymptotic expansion solves the
equations of motion, while being general enough to match the curves
arising from the integration starting at  the horizon
\bea
&& f=h_0^2+\frac{f_4}{r^4}+\frac{f^l_4}{r^4} \log r+...\;, \qquad\quad\;\;
h=h_0+{h_2 \over r^2}+{h_4 \over r^4}+{h^l_4 \over r^4}\log r+...\;,
\label{falloff1}\\
&& B=B_0+\frac{B_4}{r^4}+\frac{B^l_4}{r^4}\log r+...\;, \qquad
\hspace{-0.15in}
\quad\;\; C=C_0+\frac{C_4}{r^4}+\frac{C^l_4}{r^4}\log
r+...\;, \label{falloff2}\\
&& A_t=A_{t,0}+\frac{A_{t,2}}{r^2}+\frac{A^l_{t,2}}{r^2}\log
r+...\;, \quad A_x=A_{x,0}+{A_{x,2} \over
r^2}+\frac{A^l_{x,2}}{r^2}\log r+...\;,\label{falloff3}\\
&&\psi=\frac{\psi_1}{r}+\frac{\psi_3}{r^3}+\frac{\psi^l_3}{r^3}\log
r+...\;.\label{falloff4}
\eea
As expected, there are relations among some of the coefficients above 
(detailed in appendix A of \cite{Arean:2010wu}).
These are such that when the non-normalizable mode
$\psi_1$ is set to zero, all the
logarithmic pieces in the expansions above vanish.
Finally, the independent parameters at the boundary can be taken to be the set
\beq (h_0,
f_4, B_0, B_4, C_0, C_4, A_{t,0}, A_{t,2}, A_{x,0}, A_{x,2}, \psi_1,
\psi_3)\,.  
\eeq 
For our solutions to become asymptotically AdS as $r\to\infty$,
we must have $B_0=h_0=1$, and $C_0=\psi_1=0$.
As explained in \cite{Arean:2010wu}, the scaling symmetries (\ref{scale1B}, \ref{scale2B})
can be used to accomplish the first two conditions, 
whereas we need to shoot for the last two. We are therefore left with 
eight independent boundary data: 
\beq
(f_4, B_4, C_4, A_{t,0}, A_{t,2},A_{x,0}, A_{x,2}, \psi_3)\,.
\eeq
Since we have instead six independent data at the horizon
\eqref{horizondata5d}, we expect to find,
if any,
a two-parameter family of solutions where the condensate (dual to $\psi_3$)
does not vanish.
We choose to parametrize those solutions in terms of the superfluid velocity, and the
ratio of the temperature over the chemical potential.
Remember that according to the AdS/CFT dictionary we identify
\be
\mu\equiv A_{t,0}\,,\qquad \langle {\cal O\rangle}\equiv \sqrt{2}\psi_3,
\ee
where $\mu$, and ${\cal O}$ are, on the dual theory, 
the chemical potential and a scalar operator (of dimension 3, and dual to $\psi$)
respectively.
Moreover, the superfluid velocity in units of the chemical potential is
given by
\be
\zeta=\frac{A_{x,0}}{A_{t,0}}\,.
\ee
The temperature of the superfluid corresponds to the black hole Hawking temperature.
For a metric as \eqref{ModTisza4} it reads
\be
T = \frac{r_H^2\, f'(r_H)}{4\pi\,L^2\,h(r_H)}\,,
\label{eq:tempsf}
\ee
which can be determined in terms of the horizon data \eqref{horizondata5d}.
After some algebra, and setting $r_H=1$ as mentioned above, we get
\beq
T = \frac{1}{4\pi}\left[h_0^H
\left({9 \over 4}+2 \cosh{\psi_0^H}-{\cosh(2\psi_0^H) \over 4}\right)-
\frac{2 (A_{t,1}^H)^2}{9 h_0^H} \right]
\label{eq:temp5d}\,.
\eeq

\noindent
\bibliographystyle{JHEP}
\bibliography{AdS_CFT}
\end{document}